\documentclass[prc,preprint,showpacs,tightenlines,%
                floatfix,amsfonts,nofootinbib]{revtex4}
\usepackage{latexsym}
\usepackage{sparticles}
\usepackage{feynmf}
\usepackage{graphicx}

\begin{document}

%
% Begin: Simple substitution macros used in the text
%
%%%  syntax:
%
%\newcommand{macro name}[arguments]{what it means}
%
%*************************************************************************
\def\psibar{\overline\psi}
\def\tr{\mathop{\rm tr}\nolimits}
\def\Tr{\mathop{\rm Tr}\nolimits}
\def\zz{\hphantom{-}}          %% use in math mode

%*************************************************************************

\title{Two-Loop Corrections for Nuclear Matter in a\\
       Covariant Effective Field Theory}

\author{Ying Hu}\email{\texttt{yihu_us@yahoo.com}}\altaffiliation[permanent
address: ]{Motorola Inc., 1501 W. Shure Dr., Mail Drop: 3--2G, Arlington Heights, IL 60004.}
\author{Jeff McIntire}\email{\texttt{jewmcint@indiana.edu}}
\author{Brian D. Serot}\email{\texttt{serot@indiana.edu}}
 \affiliation{Department of Physics and Nuclear Theory Center
             Indiana University, Bloomington, IN\ \ 47405}

%%%%  The following lines FOOL REVTeX_4 into leaving some space
%     before the date in preprint mode.
%
\author{\null}
\noaffiliation

% use this for the draft version
%\date{\today ;\ \ {\bf DRAFT}}
%
% use one of these for the final version
\date{\today\\[20pt]}
%\date{May, 2007}

\begin{abstract}
%\hfill\\[-20pt]     % for more space with tightenlines

Although one-loop calculations provide a realistic description of
bulk and single-particle nuclear properties, it is necessary to
examine loop corrections to develop a systematic finite-density
power-counting scheme for the nuclear many-body problem when loops
are included.  Moreover, it is imperative to study exchange and
correlation corrections systematically to make reliable predictions
for other nuclear observables. One must also verify that the natural
sizes of the one-loop parameters are not destroyed by explicit
inclusion of many-body corrections. The loop expansion is applied to
a chiral effective hadronic lagrangian; with the techniques of
Infrared Regularization, it is possible to separate out the
short-range contributions and to write them as local products of
fields that are already present in our lagrangian. (The appropriate
field variables must be re-defined at each order in loops.) The
corresponding parameters implicitly include short-range effects to
all orders in the interaction, so these effects need not be
calculated explicitly. The remaining (long-range) contributions that
must be calculated are nonlocal and resemble those in conventional
nuclear-structure calculations.  Calculations at the two-loop level
are carried out to illustrate these techniques at finite densities
and to verify that the coupling parameters remain natural when
fitted to the empirical properties of equilibrium nuclear matter.

\end{abstract}

\smallskip

\maketitle

\section{Introduction}
\label{sec:intro}

Quantum Hadrodynamics (QHD) is a low-energy effective theory of the
strong interaction that describes the strong nuclear force by the
exchange of mesons between nucleons (the observed degrees of freedom
at this energy scale). Initially proposed by Walecka
\cite{ref:Wa74}, it has evolved over the years into a framework
based on effective field theory (EFT) and density functional theory
(DFT)
\cite{ref:Fu97,ref:Se97,ref:Fu99,ref:Fu03,ref:Se04,ref:Fu04,ref:Wa04}.
EFT embodies basic principles that are common to many areas of
physics, such as the separation of length scales in the description
of natural phenomena.  In EFT, the long-range (nonlocal) dynamics is included
explicitly, while the short-range (local) dynamics is parametrized
generically; all of the dynamics is constrained by the symmetries of
the interaction. DFT tells us that the nuclear many-body system can
be described by a universal energy functional that depends on
nuclear densities and four-vector currents. With knowledge of the
energy functional, one can calculate any observable for the
(zero-temperature) many-body system. Moreover, a simplified
treatment of the functional based on quasi-particle orbitals still
provides an exact description of the bulk properties and some
single-particle observables. Thus knowledge of the many-particle
wave function is not needed to calculate this subset of observables
\cite{ref:Se04}. The energy functional is constructed as an
expansion in small parameters (the mean meson fields, which are in
fact Kohn--Sham potentials, divided by a heavy mass, which
could be the nucleon mass or the chiral symmetry breaking scale)
and includes all possible terms consistent with the underlying
symmetries of the system.

In the current formalism, field redefinitions are employed to place
the complexity of the problem in the meson field self-interactions.
Each term is characterized by an unknown coefficient which, once all
the dimensional and combinatorial factors have been removed, is a
dimensionless constant of order unity, an assumption known as
\textit{naturalness} \cite{ref:Ma84,ref:Ge93}. The natural
separation of length scales of the system is embodied in this
theory: the long-range dynamics is included explicitly and the
short-range physics is contained in the parametrization. While this
theory in principle contains all possible terms consistent with the
underlying symmetries of QCD, in practice this is a perturbative
expansion for the energy functional
that can be truncated at a manageable level. Once it has
been truncated, the now finite number of coefficients are fixed by
experimental data; this theory can then be used for predictive
purposes \cite{ref:Mc02,ref:Mc04,ref:Mc05,ref:He02}.

QHD is a strong-coupling theory. Unlike QCD, QHD is not
asymptotically free, and currently no lattice version exists. Moreover, unlike QED,
the QHD couplings are large, and there is no obvious asymptotic
expansion to use to obtain results and refine them systematically.
It is thus unknown whether QHD permits any expansion for systematic
computation and refinement of theoretical results. One possibility
is the loop expansion, which was partially explored in
\cite{ref:Fu89,ref:Ta95}. In that work, explicit calculation of the
short-range dynamics in terms of nucleons and heavy mesons
produced enormous contributions to the energy, which rendered
the loop expansion useless. This problem was only partially avoided
by the inclusion of vertex corrections in the loops. In contrast,
with the ideas of EFT, we can now provide a straightforward, physically 
motivated discussion of the short-range dynamics.

In the present work, the loop expansion for QHD is constructed in the usual
manner \cite{ref:Di33,ref:It80,ref:Se86,ref:Fu89}. First we define
the action and the exact ground-state generating functional through
a path integral. This generating functional contains all possible
diagrams. The path integral is used to define the effective action,
which is then expanded in powers of $\hbar$ (where $\hbar$ acts as a
bookkeeping parameter and is not necessarily small
\cite{ref:Co73,ref:Il75,ref:Co77}). An equivalent way to state this
is that the effective action is expanded around its classical value
by grouping terms according to the number of quantum loops in their
corresponding diagrams. Then we perform the functional derivatives
and acquire the loop integrals. For the purposes of this work, we
are interested in the expansion only up to the two-loop level.
Consideration of the effects of contributions from the three-loop
level and higher will be considered in future investigations \cite{ref:Mc07,ref:Mc07b}. All of
the integrals that represent tadpole and disconnected diagrams
cancel out in the two-loop effective action, and we are left with only the fully connected diagrams.
In the nuclear matter limit, the effective action is proportional to
the energy density. For the cases considered here, there are three
integrals of interest. These three integrals each have two factors
of the nucleon propagator and one meson propagator (either scalar,
vector, or pion).

Why a loop expansion? The loop expansion is a simple and
well-developed expansion scheme in powers of $\hbar$ that is derived
from the path integral. The mean meson fields are included
non-perturbatively and the correlations are included perturbatively.
Therefore, one can analyze the many-body effects order by order. 
Indeed, previous work has shown the importance of ``Hartree dominance''
\cite{ref:Fu96}: the mean-field terms dominate the nuclear energy,
and exchange and correlation effects do not significantly modify the energy
or nucleon self-energies, at least for states in the Fermi sea.
We stress that we are not certain that the loop expansion is practical;
the answer to this question is left for future consideration.
However, the loop expansion has the advantage that it is fairly easy
to separate the short-range and long-range dynamics and to analyze
their structures.

The nucleon propagator can be separated into two components, known
as the Feynman and Density parts \cite{ref:Se86}. This is
accomplished by taking into account the proper pole structure of the
propagator. The Feynman part describes the propagation of a baryon
or antibaryon; the Density part involves only on-shell propagation
in the Fermi sea and incorporates the exclusion principle.

As a result, the two-loop integrals can each be separated into three
distinct parts, which we refer to as exchange, Lamb-shift, and
vacuum-fluctuation contributions \cite{ref:Fu89}. The exchange term
has two factors of the Density portion of the nucleon
propagator. Thus both momentum integrals are entirely within the
Fermi surface and the exchange term is finite. This represents a contribution from
long-range (nonlocal) physics and must be calculated explicitly. The
Lamb-shift term contains both Feynman and Density parts of
the nucleon propagator. This term is so named because it is
analogous to the Lamb shift in atomic physics, where a particle in
an occupied state interacts with a virtual particle in an unoccupied
state that shifts its spectrum. This contribution is short-range,
and we will show that it can be expressed as a sum of terms that are
already present in the QHD EFT lagrangian. As this is an effective
theory, the coefficients of these terms are determined by matching
to empirical data; thus, these short-range terms are just absorbed
into local terms already present in the lagrangian and should not be
calculated explicitly. The vacuum fluctuation term contains
two factors of the Feynman propagator and both momentum
integrals extend outside the Fermi sphere. This involves the
excitation of $N\overline{N}$ pairs and is therefore also
short-range physics. We will show that it can also be expressed as a
sum of terms which exist in the EFT lagrangian. As a result, it can
be removed in the same manner as the Lamb shift. In addition, if
nonlinear meson self-interactions are included, a number of pure
meson loops arise. These terms, however, can be expressed as a power
series in the meson fields with undetermined coefficients. As
before, these terms are just absorbed and do not need to be
calculated explicitly. The result is that, for the cases considered
here, only three finite integrals representing the two-loop
contributions from the scalar and vector meson and the pion need to
be calculated.

It is interesting to note that the procedure described above is
similar to Infrared Regularization. In Infrared
Regularization, the one-loop self-energy contribution can be
separated into soft and hard parts
\cite{ref:Ta96,ref:El98,ref:Be99,ref:Be00,ref:Sc03} (they are
sometimes referred to as the infrared singular and regular parts
\cite{ref:Ta96,ref:El98}). The hard (or infrared regular) parts
arise from large momentum, or high-energy, contributions and are
expressible as a power series of terms already contained in the
underlying lagrangian; as in our case, they are just absorbed into
the coefficients. The soft (or infrared singular) parts, in the
notation of Ellis and Tang, contain both analytic and nonanalytic
terms. The analytic portion results from high-energy dynamics and is
therefore treated in the same manner as the hard contributions. The
nonanalytic portion develops from low momentum, or low-energy
physics, and is essentially long-range dynamics. This is the
nonlocal contribution that must be calculated explicitly.
The regularization procedure for the (closed) energy loops in this
work is significantly simpler than the procedure for diagrams with 
external momenta, like scattering amplitudes or self-energies,
because the momenta carried by the boson propagators are always spacelike.

In the QHD lagrangian, a well-developed mean-field power-counting
scheme has been devised \cite{ref:Fu97,ref:Se97}. As noted, the
energy functional is an expansion in small parameters (the meson
fields divided by the nucleon mass). In addition, the Fermi wave
number, which is related to the size of the derivatives, is also
small when divided by the nucleon mass. If naturalness holds, then
adding up the powers of these ratios (and some counting factors)
yields an accurate estimate of the size of a given term. This paper
will seek to investigate the relationship between the two-loop
integrals and this underlying power counting scheme.

The purpose of this work is to illustrate these techniques by
working to the two-loop level in a loop expansion; the resulting
two-loop integrals are separated into long-range and short-range
physics. The short-range contributions are expressed in forms that
already appear at the mean field level. Since the coefficients of
these terms have yet to be determined, they are just redefined,
thereby incorporating these new contributions into the mean field
lagrangian. As a result, they are already present in the one-loop
level QHD calculation. The long-range, nonlocal physics must be
explicitly calculated. In this work, we fit the two-loop energy to the
equilibrium point of nuclear matter. Then we compare sets
constructed at the two-loop level with those of previous work
developed at the mean field level and consider the naturalness of
the parameters.  We also examine how these new contributions fit
into the power counting scheme. Some of this work has been discussed
previously in an unpublished Ph.D. dissertation \cite{ref:Hu00}.

\section{Theory}

In this section, we follow the usual procedure for constructing the
loop expansion \cite{ref:Se86} and in the following subsections
examine the one- and two-loop contributions, as in \cite{ref:Fu89}.

\subsection{Loop Expansion---Background}

Consider the following nonrenormalizable effective lagrangian, which
extends the Walecka model to include the nonrenormalizable $\pi N$
coupling and nonlinear scalar and vector meson interactions. This
lagrangian can also be obtained from the \textit{chirally invariant}
lagrangian in \cite{ref:Fu97} by retaining only the lowest-order
terms in the pion fields---sufficient for the two-loop calculations
in this work---and a subset of the meson nonlinearities\footnote{We 
are not making a chiral expansion in powers of the pion mass.
We have simply included a pion mass for kinematical purposes in Eq.\ (\ref{eqn:lagrangian}).}:

\begin{eqnarray} {\cal L} & = &
-{\psibar}\left[\gamma_{\mu}\left(\partial_{\mu} -
ig_{V}V_{\mu}\right) - i\frac{g_{A}}{f_{\pi}}\gamma_{\mu}\gamma_{5}
\partial_{\mu}\underline{\pi} + \left(M-g_{s}\phi\right)\right]\psi \nonumber \\
& & - \frac{1}{2}\left(\partial_{\mu}\phi\right)^{2} - \frac{1}{2}m_{S}^{2}\phi^{2}
- \frac{1}{4}V_{\mu\nu}V_{\mu\nu} - \frac{1}{2}m_{V}^{2}V_{\mu}V_{\mu}
- \frac{1}{2}\left(\partial_{\mu}\pi_{a}\right)^{2} - \frac{1}{2}m_{\pi}^{2}\pi_{a}^{2} \nonumber
\\[5pt]
& & {} +{\cal L}_{NL} + \delta{\cal L} \ , \label{eqn:lagrangian}
\end{eqnarray}

\noindent where
$V_{\mu\nu}=\partial_{\mu}V_{\nu}-\partial_{\nu}V_{\mu}$ and
$\underline{\pi}=\frac{1}{2}\pi_{a}\cdot\tau_{a}$. Here $\psi$ are
the fermion fields and $\phi$, $V_{\mu}$, and $\pi_{a}$ are the
meson fields (isoscalar-scalar, isoscalar-vector, and
isovector-pseudoscalar, respectively) and the heavy meson fields
are also chiral scalars. For the purposes of this
work, we will not include effects from the isovector-vector channel
or the electromagnetic field, but to do so is straightforward
\cite{ref:Fu97,ref:Se07}. The numerically small tensor coupling
between the omega meson and the nucleon is not included, and
nonlinearities in the meson sector of the lagrangian (${\cal L}_{NL}$) will be
considered later at the mean field level; the study of their effects
at two-loop order is left for future work \cite{ref:Mc07}.
$\delta{\cal L}$ contains all of the counterterms. Note that in this
work, the conventions of \cite{ref:Wa04} are used.

The action is defined as
\begin{equation}
S[\phi,V_{\mu}]\equiv\int d^{4}x \,{\cal L}(x)\ ,
\end{equation}

\noindent and the exact ground-state to ground-state generating
functional is defined through the following path integral \cite{ref:Fu89}
\begin{eqnarray}
Z[j,J_{\mu}] & \equiv & \exp\left\{iW[j,J_{\mu}]/\hbar\right\} \nonumber \\
& = & {\cal N}^{-1}\int D({\psibar})D(\psi)D(\phi)D(V_{\mu})D(\pi_{a}) \nonumber \\
& & \times \exp\left\{\frac{i}{\hbar} \int d^{4}x\left[{\cal L}(x) +
j(x)\phi(x) + J_{\mu}(x)V_{\mu}(x)\right]\right\}\ , \label{eqn:3}
\end{eqnarray}

\noindent where
\begin{equation}
{\cal N} \equiv \int D({\psibar})D(\psi)D(\phi)D(V_{\mu})D(\pi_{a})
\exp\left\{\frac{i}{\hbar} \int d^{4}x \,{\cal L}(x)\right\}\ .
\end{equation}

\noindent Here $\cal N$ is the normalization factor (in effect, the
vacuum subtraction), and $j(x)$ and $J_{\mu}(x)$ are the external
sources corresponding to the meson fields $\phi$ and $V_{\mu}$,
respectively. The connected generating functional $W[j,J_{\mu}]$
contains all of the connected diagrams, and $Z[j,J_{\mu}]$ contains
all possible diagrams without ``vacuum bubbles''.

The classical values of the meson fields are determined by
extremizing the action:
\begin{eqnarray}
\left[\frac{\delta S}{\delta \phi(x)}\right]_{\phi=\phi_{0}}=
\left(\partial^{2}-m_{S}^{2}\right)\phi_{0} = -j(x) \ ,
\label{eqn:CC1} \\[5pt]
\left[\frac{\delta S}{\delta V_{\mu}(x)}\right]_{V_{\mu}=
V_{\mu}^{0}}= -\partial_{\nu}V_{\mu\nu}^{0}-m_{V}^{2}V_{\mu}^{0} =
-J_{\mu}(x) \ . \label{eqn:CC2}
\end{eqnarray}

\noindent Next, we replace the fields by quantum fluctuations around
their classical fields: ${\psibar}(x) \rightarrow
\hbar^{1/2}{\psibar}(x)$, $\psi(x) \rightarrow \hbar^{1/2}\psi(x)$,
$\phi(x) \rightarrow \phi_{0}(x) + \hbar^{1/2}\sigma(x)$,
$V_{\mu}(x) \rightarrow V_{\mu}^{0}(x) + \hbar^{1/2}\tilde{\eta}_{\mu}(x)$,
and $\pi_{a}(x) \rightarrow \hbar^{1/2}\Omega_{a}(x)$. Notice that
the fermion fields have no classical limit (and the pion has no mean
field if one assumes no pion condensate). Observe that a factor of
$\hbar^{1/2}$ is associated with each quantum fluctuation; the
numerical value of $\hbar$ is immaterial, as it is just a
bookkeeping parameter \cite{ref:Co73,ref:Il75,ref:Co77}. Next, a
number of extra sources are included that are set equal to zero at
the end ($u$, $U_{\mu}$, $\zeta_{a}$, $\xi$, and $\bar{\xi}$
corresponding to the quantum fluctuations $\sigma$, $\tilde{\eta}_{\mu}$,
$\Omega_{a}$, ${\psibar}$, and $\psi$, respectively). The path
integral is then rewritten in terms of functional derivatives with
respect to these new sources. The final result is \cite{ref:Fu89}
\begin{eqnarray}
\lefteqn{Z[j,J_{\mu}] = {\cal N'}^{-1} \exp\left\{\frac{i}{\hbar}
\int d^{4}x\left[{\cal L}_{0}(x)+j(x)\phi_{0}(x)+J_{\mu}(x)V_{\mu}^{0}(x)
\right]\right\}} & & \nonumber \\
& & \times \exp\left\{\tr \;
\ln\left[G_{F}^{0}G_{H}^{-1}\right]\right\}\left[\!\!\left[\exp\left\{i\hbar^{1/2}
\int d^{4}x \left[\frac{i\delta}{\delta
\xi(x)}\right]\left(ig_{V}\gamma_{\mu}\left[\frac{-i\delta}{\delta
U_{\mu}(x)}\right] \right.\right.\right.\right. \nonumber \\ & &
\left.\left. {} +
i\frac{g_{A}}{f_{\pi}}\left(\gamma_{\mu}\gamma_{5}\partial_{\mu}
\left[\frac{-i\delta}{\delta \zeta_{a}(x)}\right] \cdot
\frac{\tau_{a}}{2}\right)
+ g_{s}\left[\frac{-i\delta}{\delta u(x)}\right]\right)\left[\frac{-i\delta}{\delta \bar{\xi}(x)}\right]\right\} \nonumber \\
& & \times \exp\left\{-i \int\int d^{4}xd^{4}y\,
\bar{\xi}(x)G_{H}(x-y)\xi(y)\right\} \nonumber \\
& & \times \exp\left\{\frac{i}{2}\int\int d^{4}xd^{4}y
\left[u(x)\Delta_{S}^{0}(x-y)u(y) \right.\right. \nonumber \\
& & \left.\left.\left.\left. {} + U_{\mu}(x){\cal
D}_{\mu\nu}^{0}(x-y)U_{\nu}(y) +
\zeta_{a}(x)\Delta_{\pi}^{ab}(x-y)\zeta_{b}(y)
\right]\vphantom{\int}\right\}\right]\!\!\right]_{\mathrm{sources}\
=\ 0}\ ,
\end{eqnarray}

\noindent where ${\cal N'}$ is equal to the portion of $Z$ involving the 
variational derivatives, but with free propagators.
${\cal L}_{0}$ represents the lagrangian at the mean field level.
The fermion propagators in momentum space are
\begin{eqnarray}
G_{F}^{0}(k) & = & \frac{-1}{i{\not\! k}+M-i\epsilon} \ , \\
G_{H}(k) & = & \frac{-1}{i{\not\!
k}-ig_{V}\gamma_{\mu}V_{\mu}^{0}+
\left(M-g_{S}\phi_{0}\right)}\ ,
\end{eqnarray}

\noindent and the free meson propagators in momentum space are
\begin{eqnarray}
\Delta_{S}^{0}(k) & = &
\frac{1}{k^{2}+m_{S}^{2}-i\epsilon} \ , \label{eqn:prop1} \\  {\cal
D}_{\mu\nu}^{0}(k) & = &
\frac{1}{k^{2}+m_{V}^{2}-i\epsilon} \left(\delta_{\mu\nu}
+ \frac{k_{\mu}k_{\nu}}{m_{V}^{2}}\right) \ , \label{eqn:prop2}\\
\Delta_{\pi}^{ab}(k) & = &
\frac{1}{k^{2}+m_{\pi}^{2}-i\epsilon}\,\delta_{ab}\ . \label{eqn:prop3}
\label{eqn:PION}
\end{eqnarray}

It is assumed that all of the divergent integrals are regularized in
some fashion that preserves the symmetries of the theory, for example,
dimensional regularization. The longitudinal term in the vector
meson propagator vanishes in the following analysis, as the vector
meson couples to the conserved baryon current \cite{ref:Se86}. As
usual, the infinitesimal $\epsilon$ is introduced to generate the
proper pole structure. 

The expectation values of the meson fields in
the presence of external sources are
\begin{equation}
\phi_{e}(x) = \frac{\langle0^{+}|\hat{\phi}(x)|0^{-}\rangle}
{\langle0^{+}|0^{-}\rangle} = -i\hbar \frac{\delta \ln
Z[j,J_{\mu}]}{\delta j(x)} = \frac{\delta W[j,J_{\mu}]}{\delta j(x)}
\ , \label{eqn:BB1}
\end{equation}
\begin{equation}
V_{\mu}^{e}(x) = \frac{\langle0^{+}|\hat{V}_{\mu}(x)|0^{-}\rangle}
{\langle0^{+}|0^{-}\rangle} = -i\hbar \frac{\delta \ln
Z[j,J_{\mu}]}{\delta J_{\mu}(x)} = \frac{\delta W[j,J_{\mu}]}{\delta
J_{\mu}(x)} \ . \label{eqn:BB2}
\end{equation}

\noindent In the limit of vanishing sources, these expectation values become
\begin{eqnarray}
\lim_{j\, \rightarrow \, 0} \phi_{e} & = & \overline{\phi} =
\mathrm{constant} \ , \\
\lim_{J_{\mu} \, \rightarrow \, 0} V_{\mu}^{e} & = &
\overline{V}_{\mu} = \mathrm{constant} \ .
\end{eqnarray}

\noindent We now define the effective action $\Gamma$ by a
functional Legendre transformation:
\begin{equation}
\Gamma[\phi_{e},V_{\mu}^{e}] \equiv W[j,J_{\mu}] - \int d^{4}x
\left[j(x)\phi_{e}(x) + J_{\mu}(x)V_{\mu}^{e}(x)\right] \ .
\end{equation}

\noindent In uniform nuclear matter, this effective action is
related to the energy density ${\cal E}$ by
\begin{equation}
\lim_{j,J_{\mu} \, \rightarrow \, 0}\Gamma[\phi_{e},V_{\mu}^{e}] = -
\int d^{4}x \,{\cal E}[\overline{\phi},\overline{V}_{\mu}] \
.\label{eqn:DD}
\end{equation}

\subsection{One-Loop Calculation}

In this section, we consider the lowest-order terms in the loop
expansion, the one-loop contributions; these are the terms that
correspond to $O(\hbar^{0})$ in the expansion. The generating
functional at one-loop order is \cite{ref:Fu89}
\begin{eqnarray}
Z^{(1)}[j,J_{\mu}] & \equiv & \exp\left\{iW^{(1)}[j,J_{\mu}]/\hbar\right\} \nonumber \\
& = & \exp\left\{\frac{i}{\hbar} \int d^{4}x\,\left[{\cal L}_{0}(x)
+ j(x)\phi_{0} + J_{\mu}(x)V_{\mu}^{0} \right]\right\} \nonumber
\\ & & \times \exp\left\{\tr \;
\ln\left[G_{F}^{0}G_{H}^{-1}\right]\right\} \ , \label{eqn:1a}
\end{eqnarray}

\noindent where we have assumed uniform classical fields. Therefore,
the connected generating functional is (for now we have dropped the
nonlinear meson self-interactions)
\begin{eqnarray}
W^{(1)}[j,J_{\mu}] & = & \int d^{4}x\,\left\{-
\frac{1}{2}m_{S}^{2}\phi_{0}^{2} -
\frac{1}{2}m_{V}^{2}V_{\mu}^{0}V_{\mu}^{0} + j\phi_{0} +
J_{\mu}V_{\mu}^{0}\right. \nonumber \\ & & \left. - i\hbar \int
\frac{d^{4}k}{(2\pi)^{4}} \tr \; \ln\left[1 -
\frac{ig_{V}\gamma_{\mu}V_{\mu}^{0} +g_{S}\phi_{0}} {i{\not\! k} + M
- i\epsilon}\right]\right\} \ .
\end{eqnarray}

\noindent Here ``tr'' denotes the summation over both spin and
isospin. Following from the previous section, Eqs.\ (\ref{eqn:BB1})
and (\ref{eqn:BB2}) become $\phi_{e}^{(1)}(x) = \phi_{0}$ and
${V_{\mu}^{e}}^{(1)}(x) = V_{\mu}^{0}$, respectively. For a
spatially uniform system, the classical vector field is
$V_{\mu}^{0}=iV_{0}\delta_{\mu 4}$. To calculate the momentum
integral, we use the relation \cite{ref:Fu89}
\begin{eqnarray}
\lefteqn{-i\hbar \int \frac{d^{4}k}{(2\pi)^{4}} \tr \; \ln\left[1 +
\frac{g_{V}\gamma_{4}V_{0} - g_{S}\phi_{0}}
{i{\not\! k} + M}\right]} & & \nonumber \\
& = & i\hbar\int \frac{d^{4}k}{(2\pi)^{4}}
\tr\left[\gamma_{4}k_{4}\left(G_{H}(k) - G_{F}^{0}(k)\right)\right]
\ ,
\end{eqnarray}

\noindent where $G_{F}^{0}$ comes from the vacuum subtraction ${\cal
N}$, which determines the proper boundary conditions. 
Then, using dimensional regularization to eliminate the vector
field in the momentum integrals, the effective action becomes
\begin{eqnarray}
\Gamma^{(1)}[\phi_{e},V_{e}^{0}] & = & \int
d^{4}x\,\left\{\frac{1}{2}m_{V}^{2}V_{0}^{2} -
\frac{1}{2}m_{S}^{2}\phi_{0}^{2} - \hbar g_{V}V_{0}\rho_{B} \right.
\nonumber \\ & &
- \hbar\,\frac{\gamma}{(2\pi)^{3}}\int d^{3}k E^{*}(k)\theta(k_{F}-|\vec{k}|)
\nonumber \\
& & \left.+ i\hbar\int \frac{d^{4}k}{(2\pi)^{4}}\,k_{4}
\tr\left[\gamma_{4}G_{F}^{*}(k)\right] - i\hbar\int
\frac{d^{4}k}{(2\pi)^{4}}\,k_{4}
\tr\left[\gamma_{4}G_{F}^{0}(k)\right]\right\}\ , 
\label{eqn:7}
\end{eqnarray}

\noindent where $E^{*}(k) \equiv (\vec{k}^{2} + {M^{*}}^{2})^{1/2}$,
the effective nucleon mass $M^{*} \equiv M - g_{S}\phi_{0}$, and
$\gamma$ is the spin--isospin degeneracy. Here we have separated the
nucleon propagator into two parts \cite{ref:Se86} (known as the
Feynman and Density contributions), as shown below:
\begin{eqnarray}
G^{*}(k) & = & \frac{-1}{i{\not\! k}+M-g_{S}\phi_{0}} \nonumber \\
& = & \left(i{\not\! k}-M^{*}\right)\left[\frac{1}{k^{2}+{M^{*}}^{2}-i\epsilon}
- \frac{i\pi}{E^{*}(k)}\delta[k_{4} - E^{*}(k)]\theta(k_{F}-|\vec{k}|)\right] \nonumber \\
& \equiv & G_{F}^{*}(k) + G_{D}^{*}(k) \ .
\label{eqn:BB3}
\end{eqnarray}

\noindent This is done by taking the proper pole structure into
account. The Feynman part describes the propagation of baryons and
antibaryons; the Density part involves only on-shell
propagation in the Fermi sea and corrects the propagation of
positive-energy baryons for the Pauli exclusion principle.

To remove the divergences in the momentum integral in
Eq.~(\ref{eqn:7}), we expand $G_{F}^{*}(k)$ as a polynomial in
$g_{S}\phi_{0}$ (the Furry expansion \cite{ref:Fu89})
\begin{equation}
G_{F}^{*}(k) = \sum_{n\, =\, 0}^{m}\left(M^{*}-M\right)^{n}
\left[G_{F}^{0}(k)\right]^{n+1}
+\left(M^{*}-M\right)^{m+1}\left[G_{F}^{0}(k)\right]^{m+1}
G_{F}^{*}(k)\ , \label{eqn:fur}
\end{equation}

\noindent which is valid for any $m \geq 0$. If we take $m=4$ and
insert into Eq.\ (\ref{eqn:7}), we notice that the first four terms
are divergent. The counterterms introduced to absorb these
divergences are
\begin{equation}
\delta{\cal L}^{(1)} = \hbar\sum_{n\, =\,
1}^{4}\alpha_{n}\phi^{n}_{0}\ , \label{eqn:10a}
\end{equation}

\noindent where $\alpha_{i}$ are pure numbers. The first term
in the Furry expansion cancels the last term in Eq.\ (\ref{eqn:7});
the next four terms cancel with the counterterms in Eq.\
(\ref{eqn:10a}). Thus, the one-loop effective action is just
\begin{eqnarray}
\Gamma^{(1)}[\phi_{e},V_{e}^{0}] & = & \int
d^{4}x\left\{\frac{1}{2}m_{V}^{2}V_{0}^{2} -
\frac{1}{2}m_{S}^{2}\phi_{0}^{2} - g_{V}V_{0}\rho_{B}\right.
\nonumber \\ & & \left. - \frac{\gamma}{(2\pi)^{3}}\int d^{3}k
E^{*}(k)\theta(k_{F}-|\vec{k}|) - \Delta{\cal
E}_{VF}(M^{*})\right\}\ ,
\end{eqnarray}

\noindent where the factor $\hbar$ has been omitted and
\begin{equation}
\Delta{\cal E}_{VF}(M^{*}) \equiv
-i\left(M^{*}-M\right)^{5}\int\frac{d^{4}k}{(2\pi)^{4}} \,k_{4}
\tr\left\{\gamma_{4}
\left[G_{F}^{0}(k)\right]^{5}G_{F}^{*}(k)\right\}\ . \label{eqn:EVF}
\end{equation}

As before, the mean fields are determined by extremizing the effective action.
At the one-loop level, the energy density is [using Eq.\
(\ref{eqn:DD})]
\begin{eqnarray}
{\cal E}^{(1)}[M^{*},\rho_{B}] & = & g_{V}V_{0}\rho_{B} -
\frac{1}{2}m_{V}^{2}V_{0}^{2} + \frac{m_{S}^{2}}{2g_{S}^{2}}
(M-M^{*})^{2} \nonumber \\ & & + \frac{\gamma}{(2\pi)^{3}}\int
d^{3}k E^{*}(k)\theta(k_{F}-|\vec{k}|) + \Delta{\cal E}_{VF}(M^{*})\
,
\end{eqnarray}

\noindent which is the relativistic Hartree approximation in the
original Walecka model \cite{ref:Wa74}. The final term in the energy
density is written as \cite{ref:Fu89}
\begin{eqnarray}
\Delta{\cal E}_{VF}(M^{*}) & = & 
\frac{M^{4}}{4\pi^{2}}\left\{ \frac{(g_{S}\phi_{0})^{5}}{5M^{5}} +
\frac{(g_{S}\phi_{0})^{6}}{30M^{6}} +
\frac{(g_{S}\phi_{0})^{7}}{105M^{7}} + \cdots \right. \nonumber \\[5pt] &
& \left. \quad {} +
\frac{4!(n-5)!}{n!}\frac{(g_{S}\phi_{0})^{n}}{M^{n}} + \cdots
\right\} \label{eqn:exp}
\end{eqnarray}

\noindent Note that this term has no explicit density dependence.

As discussed in \cite{ref:Fu97a,ref:Se97}, the general form of each
term in $\Delta{\cal E}_{VF}$ shows that this vacuum contribution is
unnaturally large. If we accept the naturalness assumption, the
conclusion is that the vacuum contribution is not well described by
$\Delta{\cal E}_{VF}$ at the one-baryon-loop level. Baryons
are incorrect degrees of freedom for computing short-range loops.
Furthermore, to make these terms natural, it is certain that there must be large
cancellations from higher-order loops, which therefore must
be calculated.  Yet, we know that in
principle the polynomial terms in $\phi$ should appear in an
effective lagrangian, since they satisfy all the symmetry
requirements; moreover, there will always be short-range
contributions to these terms that we cannot calculate.  So the
proposal is that we need not work hard to get the vacuum
contributions---instead, we can adjust the unknown natural
coefficients to include them!

\subsection{Two-Loop Calculation}

In this section, we present the corrections to the theory arising from the two-loop contributions.
We define the connected generating functional at the two-loop level as
\begin{equation}
W^{(2)} = W^{(1)} + W_{2} \ .
\end{equation}

\noindent Keeping only the terms of $O(\hbar)$ in the loop expansion
(which is essentially an expansion in the coupling constants
$g_{S}$, $g_{V}$, and $g_{A}$), the connected generating functional
is
\begin{eqnarray}
W_{2} & = & \frac{i\hbar^{2}}{2} \int\int
d^{4}xd^{4}y\left\{g_{S}^{2}\,
\delta_{\alpha\beta}\delta_{\alpha'\beta'}
\left[\frac{-i\delta}{\delta u(x)}\right]\left[\frac{-i\delta}{\delta u(y)}\right] \right. \nonumber \\
& & {} -
g_{V}^{2}\,(\gamma_{\mu})_{\alpha\beta}(\gamma_{\nu})_{\alpha'\beta'}
\left[\frac{-i\delta}{\delta U_{\mu}(x)}\right]\left[\frac{-i\delta}{\delta U_{\nu}(y)}\right] \nonumber \\
& & \left. {} - \frac{g_{A}^{2}}{f_{\pi}^{2}}
\left(\gamma_{\mu}\gamma_{5}\partial_{\mu}^{x}\left[\frac{-i\delta}{\delta
\zeta_{a}(x)}\right] \cdot \frac{\tau_{a}}{2}\right)_{\alpha\beta}
\left(\gamma_{\nu}\gamma_{5}\partial_{\nu}^{y}\left[\frac{-i\delta}{\delta
\zeta_{b}(y)}\right]
\cdot \frac{\tau_{b}}{2}\right)_{\alpha'\beta'} \right\} \nonumber \\
& & \times \left[\frac{i\delta}{\delta \xi(x)}\right]_{\alpha}\left[\frac{-i\delta}{\delta \bar{\xi}(x)}\right]_{\beta}
\left[\frac{i\delta}{\delta \xi(y)}\right]_{\alpha'}\left[\frac{-i\delta}{\delta \bar{\xi}(y)}\right]_{\beta'} \nonumber \\
& & \times \exp\left\{-i \int\int d^{4}xd^{4}y\,
\bar{\xi}(x)G_{H}(x-y)\xi(y)\right\} \nonumber \\ & &
\times \exp\left\{\frac{i}{2}\int\int d^{4}xd^{4}y\,\left[u(x)\Delta_{S}^{0}(x-y)u(y) \right.\right. \nonumber \\
& & \left.\left.\left. {} + U_{\sigma}(x){\cal
D}_{\sigma\rho}^{0}(x-y)U_{\rho}(y) +
\zeta_{c}(x)\Delta_{\pi}^{cd}(x-y)\zeta_{d}(y)
\right]\vphantom{\int}\right\}\right|_{\mathrm{sources}\ =\ 0} \nonumber \\
& & {} - \mathrm{VEV} \ , \label{eqn:2}
\end{eqnarray}

\noindent where VEV is the vacuum subtraction, which is just
equivalent to the rest of $W_{2}$ with free propagators. After
working out the variational derivatives, the full expression becomes
\begin{eqnarray}
W_{2} & = & \frac{\hbar^{2}}{2} \int d^{4}x\int\int\frac{d^{4}k}{(2\pi)^{4}}
\frac{d^{4}q}{(2\pi)^{4}} \nonumber \\
& & \times \left(\vphantom{\int} g_{S}^{2}\left\{\Delta_{S}^{0}(k-q)
\tr\left[G_{H}(k)G_{H}(q)\right] -\Delta_{S}^{0}(0)
\tr\left[G_{H}(k)\right] \tr\left[G_{H}(q)\right]\right\}
\right. \nonumber \\[5pt]
& & \quad {} - g_{V}^{2}\left\{{\cal D}_{\mu\nu}^{0}(k-q)
\tr\left[\gamma_{\mu}G_{H}(k)\gamma_{\nu}G_{H}(q)\right] 
-{\cal D}_{\mu\nu}^{0}(0)
\tr\left[\gamma_{\mu}G_{H}(k)\right]
\tr\left[\gamma_{\nu}G_{H}(q)\right]\right\} \nonumber \\[5pt]
& & \left. \quad {} -\frac{g_{A}^{2}}{f_{\pi}^{2}}
\Delta_{\pi}^{ab}(k-q) \tr\left[(\not\! k - \not\!
q)\gamma_{5}\frac{\tau_{a}}{2}G_{H}(k) (\not\! k - \not\!
q)\gamma_{5}\frac{\tau_{b}}{2}G_{H}(q)\right] \right) \nonumber
\\[5pt]
& & {} - \mathrm{VEV} \ .
\end{eqnarray}

$W_{2}$ contains all the connected two-loop diagrams. To isolate and
remove the tadpole diagrams, we must consider the effective action.
The expectation values of the meson fields at the two-loop level are
\cite{ref:Fu89}
\begin{eqnarray}
\phi_{e}^{(2)}(x) & = & \frac{\delta W^{(1)}}{\delta j(x)} =
\phi_{0}(x) - \hbar\, g_{S}\int d^{4}y\int\frac{d^{4}k}{(2\pi)^{4}} \,\Delta_{S}^{0}(x-y)
\tr\left[G_{H}(k)\right] \ , \nonumber \\
{V_{\mu}^{e}}^{(2)}(x) & = & \frac{\delta W^{(1)}}{\delta
J_{\mu}(x)} = V_{\mu}^{0}(x) - i\hbar\, g_{V} \int d^{4}y\int\frac{d^{4}k}{(2\pi)^{4}} \,{\cal
D}_{\mu\nu}^{0}(x-y) \tr\left[\gamma_{\mu}G_{H}(k)\right] \ , 
\end{eqnarray}

\noindent where $W^{(1)}=S_{0}+W_{1}$, and there are no pion tadpoles.
Thus, the effective meson fields contain tadpole contributions.

The effective action at the two-loop level is
\begin{eqnarray}
\Gamma^{(2)}[\phi_{e},V_{\mu}^{e}] & = & S[\phi_{0},V_{\mu}^{0}]
+ W_{1}[\phi_{0},V_{\mu}^{0}] + W_{2}[\phi_{0},V_{\mu}^{0}] \nonumber \\
& & + \int d^{4}x\left\{j(x)\left[\phi_{0}(x)-\phi_{e}(x)\right]
+J_{\mu}(x)\left[V_{\mu}^{0}(x)-V_{\mu}^{e}(x)\right]\right\} \nonumber \\
& & {}+ O(\hbar^{3})\ .
\end{eqnarray}

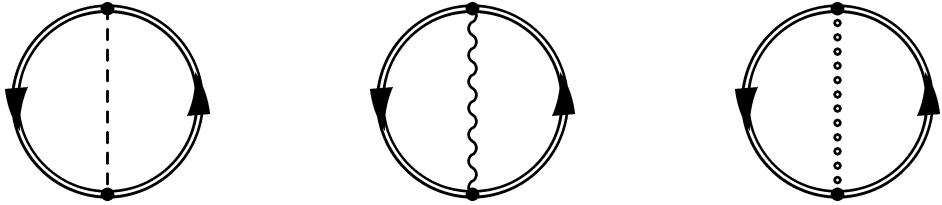
\begin{figure}[!htb]
  \begin{center}
    \begin{tabular}{ccc}
      \begin{fmffile}{fd1}
    \fmfframe(1,2)(1,2){
      \begin{fmfgraph}(70,70)
        \fmfi{dbl_plain_arrow}{.5[nw,ne] .. .5[nw,sw] .. .5[sw,se]}
        \fmfi{dbl_plain_arrow}{.5[sw,se] .. .5[ne,se] .. .5[nw,ne]}
        \fmfi{dashes}{.5[nw,ne] .. .5[sw,se]}
        \fmfiv{d.sh=circle,d.siz=2thick}{.5[nw,ne]}
        \fmfiv{d.sh=circle,d.siz=2thick}{.5[sw,se]}
      \end{fmfgraph}
    }
      \end{fmffile}

      &\qquad\qquad

      \begin{fmffile}{fd2}
    \fmfframe(1,2)(1,2){
      \begin{fmfgraph}(70,70)
        \fmfi{dbl_plain_arrow}{.5[nw,ne] .. .5[nw,sw] .. .5[sw,se]}
        \fmfi{dbl_plain_arrow}{.5[sw,se] .. .5[ne,se] .. .5[nw,ne]}
        \fmfi{photon}{.5[nw,ne] .. .5[sw,se]}
        \fmfiv{d.sh=circle,d.siz=2thick}{.5[nw,ne]}
        \fmfiv{d.sh=circle,d.siz=2thick}{.5[sw,se]}
      \end{fmfgraph}
    }
      \end{fmffile}

      &\qquad\qquad

      \begin{fmffile}{fd3}
    \fmfframe(1,2)(1,2){
      \begin{fmfgraph}(70,70)
        \fmfi{dbl_plain_arrow}{.5[nw,ne] .. .5[nw,sw] .. .5[sw,se]}
        \fmfi{dbl_plain_arrow}{.5[sw,se] .. .5[ne,se] .. .5[nw,ne]}
        \fmfi{dbl_dots}{.5[nw,ne] .. .5[sw,se]}
        \fmfiv{d.sh=circle,d.siz=2thick}{.5[nw,ne]}
        \fmfiv{d.sh=circle,d.siz=2thick}{.5[sw,se]}
      \end{fmfgraph}
    }
      \end{fmffile}
    \end{tabular}
    \caption{Two-loop diagrams. The double lines represent baryon propagators
      [Eq.\ (\ref{eqn:BB3})]. The dashed, wiggly, and dotted lines represent scalar, vector, and
    pion propagators, respectively [Eqs.\ (\ref{eqn:prop1}) to (\ref{eqn:prop3})].}
    \label{fig:fd}
  \end{center}
\end{figure}

\noindent Next, we change variables to $\phi_{0} = \phi_{e} +
\phi_{1}$ and $V_{\mu}^{0} = V_{\mu}^{e} + V_{\mu}^{1}$, where
$\phi_{1}$ and $V_{\mu}^{1}$ cancel quantum corrections contained in
$\phi_{e}$ and $V_{\mu}^{e}$ respectively. Then, we expand this
expression as a power series in $\phi_{1}$ and $V_{\mu}^{1}$ about
$\phi_{e}$ and $V_{\mu}^{e}$. Using Eqs.\ (\ref{eqn:CC1}) and
(\ref{eqn:CC2}), we get
\begin{eqnarray}
\Gamma^{(2)}[\phi_{e},V_{\mu}^{e}] 
& = & \Gamma^{(1)}[\phi_{e},V_{\mu}^{e}]
+ W_{2}[\phi_{e},V_{\mu}^{e}] \nonumber \\
& & {} + \frac{\hbar^{2}}{2} g_{S}^{2} \int d^{4}x\int
\frac{d^{4}k}{(2\pi)^{4}}\frac{d^{4}q}{(2\pi)^{4}}\,
\Delta_{S}^{0}(0) \tr\left[G_{H}(k)\right]
\tr\left[G_{H}(q)\right] \nonumber \\
& & {} - \frac{\hbar^{2}}{2} g_{V}^{2} \int d^{4}x\int
\frac{d^{4}k}{(2\pi)^{4}}\frac{d^{4}q}{(2\pi)^{4}} \, {\cal
D}_{\mu\nu}^{0}(0) \tr\left[\gamma_{\mu}G_{H}(k)\right]
\tr\left[\gamma_{\nu}G_{H}(q)\right] \ , 
\end{eqnarray}

\noindent where the last two terms cancel with terms in $W_{2}$ (the
tadpole diagrams drop out). Now we use dimensional regularization to
eliminate the dependence on $V_{\mu}^{0}$. Lastly, we write the
energy density as ${\cal E}^{(2)} = {\cal E}^{(1)} + {\cal E}_{2}$,
where \cite{ref:Fu89}
\begin{eqnarray}
{\cal E}_{2} & = &
-\int\int\frac{d^{4}k}{(2\pi)^{4}}\frac{d^{4}q}{(2\pi)^{4}}
\left[\frac{g_{S}^{2}}{2}\, \Delta_{S}^{0}(k-q)\left\{
\tr\left[G^{*}(k)G^{*}(q)\right]
- \tr\left[G^{0}_{F}(k)G^{0}_{F}(q)\right]\right\} \right. \nonumber \\
& & {} - \frac{g_{V}^{2}}{2}\, {\cal D}_{\mu\nu}^{0}(k-q)\left\{
\tr\left[\gamma_{\mu}G^{*}(k)\gamma_{\nu}G^{*}(q)\right]
- \tr\left[\gamma_{\mu}G^{0}_{F}(k)\gamma_{\nu}G^{0}_{F}(q)\right]\right\}
\nonumber \\
& & {} -
\frac{g_{A}^{2}}{2f_{\pi}^{2}}
\Delta_{\pi}^{ab}(k-q)\left\{ \tr\left[(\not\! k - \not\!
q)\gamma_{5}\frac{\tau_{a}}{2}G^{*}(k)
(\not\! k - \not\! q)\gamma_{5}\frac{\tau_{b}}{2}G^{*}(q)\right] \right. \nonumber \\
& & \left.\left. \qquad {} - \tr\left[(\not\! k - \not\!
q)\gamma_{5}\frac{\tau_{a}}{2}G^{0}_{F}(k) (\not\! k - \not\!
q)\gamma_{5}\frac{\tau_{b}}{2}G^{0}_{F}(q)\right]\right\} \vphantom{\int} \right]\ ,
\end{eqnarray}

\noindent and the factors of $\hbar$ have been suppressed. 
The terms involving the free nucleon propagators come from the VEV subtraction.
The corresponding two-loop diagrams are shown in Fig.~\ref{fig:fd}.

Since each $G^{*}$ can be separated into Feynman and Density parts
[see Eq.(\ref{eqn:BB3})], we can rewrite the first term in ${\cal
E}_{2}$ as the following sum \cite{ref:Fu89}
\begin{equation}
{\cal E}_{\phi}^{(2)} = {\cal E}_{\phi-EX}^{(2)} + {\cal
E}_{\phi-LS}^{(2)} + {\cal E}_{\phi-VF}^{(2)}\ ,
\end{equation}

\noindent where
\begin{eqnarray}
{\cal E}_{\phi-EX}^{(2)} & = &
-\frac{g_{S}^{2}}{2}\int\int\frac{d^{4}k}{(2\pi)^{4}}\frac{d^{4}q}{(2\pi)^{4}}\,
\Delta_{S}^{0}(k-q) \tr\left[G^{*}_{D}(k)G^{*}_{D}(q)\right]\ , \nonumber \\
{\cal E}_{\phi-LS}^{(2)} & = &
-g_{S}^{2}\int\int\frac{d^{4}k}{(2\pi)^{4}}\frac{d^{4}q}{(2\pi)^{4}}\,
\Delta_{S}^{0}(k-q) \tr\left[G^{*}_{F}(k)G^{*}_{D}(q)\right]\ , \nonumber \\
{\cal E}_{\phi-VF}^{(2)} & = &
-\frac{g_{S}^{2}}{2}\int\int\frac{d^{4}k}{(2\pi)^{4}}\frac{d^{4}q}{(2\pi)^{4}}\,
\Delta_{S}^{0}(k-q) \nonumber \\ & & \qquad\qquad \times \left\{
\tr\left[G^{*}_{F}(k)G^{*}_{F}(q)\right] -
\tr\left[G^{0}_{F}(k)G^{0}_{F}(q)\right]\right\}\ .
\end{eqnarray}

\noindent It turns out this is also true for the vector and pion
terms, so that we may write the energy density at the two-loop level
as
\begin{equation}
{\cal E}^{(2)} = {\cal E}^{(1)} + {\cal E}_{EX}^{(2)} + {\cal
E}_{LS}^{(2)} + {\cal E}_{VF}^{(2)}\ .
\label{eqn:full}
\end{equation}

${\cal E}_{EX}^{(2)}$ is called the ``exchange term''. It has two
factors of $G_{D}^{*}$, which restricts the double integral to the
inside of the Fermi sphere. It is finite and explicitly
density-dependent; therefore, we regard it as a purely many-body
effect. It can be calculated straightforwardly. This corresponds to
the exchange of identical fermions in occupied states.

${\cal E}_{LS}^{(2)}$ is analogous to the Lamb shift in atomic
physics, since it involves particles in occupied states whose
energies are shifted by interactions with the fluctuating meson
fields at finite density. These fluctuations modify the short-range
structure of the baryon due to the existence of the background meson
fields.

${\cal E}_{VF}^{(2)}$ is a true vacuum fluctuation, since it
involves both meson and baryon virtual excitations. This involves
the excitations of $N\overline{N}$ pairs, which is also short-range
physics.

The ${\cal E}_{EX}^{(2)}$ contributions contain long-range dynamics
[they are characterized by the length scale of $O(k_{F}^{-1})
\approx 0.8\, \mathrm{fm}$] and are nonlocal (they involve
logarithmic functions of momenta); thus, we must calculate them
explicitly. The contributions ${\cal E}_{LS}^{(2)}$ and ${\cal
E}_{VF}^{(2)}$ correspond to short-range physics [they are
characterized by length scales of $O(M^{-1}) \approx
0.2\,\mathrm{fm}$]. So we expect that the latter two can be absorbed
into the coefficients of the local terms in the lagrangian.

Normally, certain counterterms are introduced at the two-loop level
to deal with the divergences in ${\cal E}_{LS}^{(2)}$ and ${\cal
E}_{VF}^{(2)}$, and the remaining finite contributions, which depend
on the renormalization conditions, can be calculated numerically
\cite{ref:Fu89}. However, for a nonrenormalizable effective
lagrangian, an infinite number of counterterms are required in
principle. Here we will argue that both ${\cal E}_{LS}^{(2)}$ and
${\cal E}_{VF}^{(2)}$ can be written in forms that are already
present in our lagrangian (before truncation). Therefore, by
adjusting the coefficients of these terms, these contributions are
completely absorbed.

\subsection{Short-range Dynamics}

Consider the Lamb-shift contribution from the scalar meson
\begin{eqnarray}
{\cal E}^{(2)}_{\phi-LS} & = &
-g_{S}^{2}\int\int\frac{d^{4}k}{(2\pi)^{4}}\frac{d^{4}q}{(2\pi)^{4}}\,
\Delta_{S}^{0}(k-q) \tr\left[G^{*}_{F}(k)G^{*}_{D}(q)\right] \nonumber \\
& = & 2 \int\int\frac{d^{4}k}{(2\pi)^{4}}\,
\tr\left[G^{*}_{D}(k)\Sigma^{*\phi}_{F}(k)\right]\ ,
\end{eqnarray}

\noindent where the self-energy is
\begin{equation}
\Sigma^{*\phi}_{F}(k) = -\frac{g_{S}^{2}}{2}
\int\frac{d^{4}q}{(2\pi)^{4}}\, \Delta_{S}^{0}(k-q)G^{*}_{F}(q)\ .
\end{equation}

\noindent Note that this expression contains \textit{no explicit
density dependence}. We substitute the Furry expansion [see Eq.\
(\ref{eqn:fur})] into the self-energy:
\begin{equation}
\Sigma^{*\phi}_{F}(k) = -\frac{g_{S}^{2}}{2} \sum_{n\, =\,
0}^{\infty}\left(M^{*}-M\right)^{n} \int\frac{d^{4}q}{(2\pi)^{4}}\,
\Delta_{S}^{0}(k-q)\left[G^{0}_{F}(q)\right]^{n+1}\ ,
\label{eqn:43}
\end{equation}

\noindent but now we have let the sum go to infinity and not cut it
off as before. This can be rewritten as a Taylor series
\begin{equation}
\Sigma^{*\phi}_{F}(k) = \sum_{n\, =\, 0}^{\infty}\frac{1}{n!} \left.
\frac{d^{n}\Sigma^{*
\phi}_{F}(k)}{{dM^{*}}^{n}}\right|_{M^{*}=M}\left(M^{*}-M\right)^{n}\
.
\end{equation}

\noindent We then expand $\Sigma^{*\phi}_{F}(k)$ as a Taylor series
around $i{\not\! k}=M$ to separate the momentum dependence, or
\begin{eqnarray}
\Sigma^{* \phi}_{F}(k) & = & \Sigma^{* \phi}_{F}(i{\not\! k}=M)
+ \left. \frac{d\Sigma^{* \phi}_{F}(k)}{d(i{\not\! k})}
\right|_{i{\not\, k}=M}(i{\not\! k} - M) 
+ \left. \frac{1}{2}\frac{d^{2}\Sigma^{*
\phi}_{F}(k)}{{d(i{\not\! k})}^{2}}\right|_{i{\not\, k}=M} (i{\not\!
k} - M)^{2} + \ldots \nonumber \\ & &
\end{eqnarray}

\noindent The self-energy now becomes
\begin{eqnarray}
\Sigma^{*\phi}_{F}(k) & = & \sum_{m\, =\, 0}^{\infty}\ \sum_{n\, =\,
0}^{\infty} \frac{1}{m!n!}\, \left. \frac{d^{m+n}
\Sigma^{*\phi}_{F}(k)}{d(i{\not\! k})^{m}d{M^{*}}^{n}}
\right|_{i{\not\, k}=M,M^{*}=M} \nonumber \\
& & \qquad \times\left(i{\not\! k}-M\right)^{m}\left(M^{*}-M\right)^{n}
\nonumber \\
& = & \sum_{m\, =\, 0}^{\infty}\ \sum_{n\, =\, 0}^{\infty}
a_{mn}^{\phi}\left(i{\not\! k}-M\right)^{m}\left(M^{*}-M\right)^{n}
\ ,
\end{eqnarray}

\noindent where $a_{mn}^{\phi}$ are just pure numbers, some of which
are finite and others infinite (these correspond to divergent
diagrams). This consequence of Lorentz covariance allows one to 
rewrite the Lamb-shift contribution to the energy density as
\begin{eqnarray}
{\cal E}_{\phi-LS}^{(2)} & = & 2\sum_{m\, =\, 0}^{\infty}\ \sum_{n\,
=\, 0}^{\infty} a_{mn}^{\phi} \int\frac{d^{4}k}{(2\pi)^{4}}
\tr\left[G_{D}^{*}(k)\left(i{\not\! k}-M\right)^{m}
\left(M^{*}-M\right)^{n}\right] \nonumber \\
& = & 2\sum_{m\, =\, 0}^{\infty}\ \sum_{n\, =\, 0}^{\infty}
a_{mn}^{\phi}\left(M^{*}-M\right)^{n} 
\int\frac{d^{4}k}{(2\pi)^{4}}\frac{i\pi}{E^{*}(k)} \nonumber \\
& & \quad \times
\delta\left[k_{4}-E^{*}(k)\right]
\theta\left(k_{F}-|\vec{k}|\right)
\tr\left[\left(i{\not\!k}-M^{*}\right)\left(i{\not\! k}-M\right)^{m}\right] \ .
\label{eqn:49}
\end{eqnarray}

In general, one can write
\begin{equation}
\left(i{\not\! k}-M\right)^{m} = f_{m}(k^{2},M)\, i{\not\! k} +
g_{m}(k^{2},M)\ ,
\label{eqn:dec}
\end{equation}

\noindent where $f_{m}$ and $g_{m}$ are polynomials in $k^{2}$ and
$M$. This lets one reduce the integral to
\begin{eqnarray}
\lefteqn{-2\gamma\int\frac{d^{4}k}{(2\pi)^{4}}\left[k^{2}f_{m}(k^{2},M)
+ M^{*}g_{m}(k^{2},M)\right]} & &  \nonumber \\
\lefteqn{\quad \times \frac{i\pi}{E^{*}(k)}\,\delta\left[k_{4}-E^{*}(k)\right]
\theta\left(k_{F}-|\vec{k}|\right)} & & \nonumber \\
& = & \frac{1}{2}\, F_{m}(M^{*},M) \frac{\gamma}{(2\pi)^{3}} \int
d^{3}k \frac{M^{*}}{E^{*}(k)}\,\theta\left(k_{F}-|\vec{k}|\right)
\nonumber \\
& = & \frac{1}{2}\, F_{m}(M^{*},M) \rho_{S}\ .
\end{eqnarray}

\noindent Here $\rho_{S}$ is the scalar baryon density and the
on-shell condition has been used. Substituting this result into Eq.\
(\ref{eqn:49}), we get
\begin{eqnarray}
{\cal E}_{\phi-LS}^{(2)} & = & \rho_{S}\sum_{m\, =\, 0}^{\infty}\
\sum_{n\, =\, 0}^{\infty} a_{mn}^{\phi} F_{m}(M^{*},M)
\left(M^{*}-M\right)^{n} \nonumber \\
& = & \rho_{S}\sum_{m\, =\, 0}^{\infty}
a_{m}^{\phi}(g_{S}\phi_{0})^{m}\ . \label{eqn:els}
\end{eqnarray}

\noindent Before the field redefinitions were conducted on the
effective lagrangian to put it in canonical form, terms like
$\rho_{S}(g_{S}\phi_{0})^{n} =
\langle{\psibar}\psi\rangle(g_{S}\phi_{0})^{n}$ appeared in the
theory. Since the coefficients of these terms will be eliminated by
field redefinitions, the two-loop contributions to these terms can
just be absorbed, and there is no need to calculate them.
These arguments rely only on the Lorentz structure of the self-energy
and the on-shell condition imposed by the Density propagator,
so similar conclusions follow for the contributions from the
vector mesons and pions.

We treat the term ${\cal E}_{\phi-VF}^{(2)}$ in a similar manner:
\begin{eqnarray}
{\cal E}^{(2)}_{\phi-VF} & = &
-\frac{g_{S}^{2}}{2}\int\int\frac{d^{4}k}{(2\pi)^{4}}
\frac{d^{4}q}{(2\pi)^{4}}\,
\Delta_{S}^{0}(k-q) 
\left\{\tr\left[G^{*}_{F}(k)G^{*}_{F}(q)\right]
- \tr\left[G^{0}_{F}(k)G^{0}_{F}(q)\right]\right\}\ . \nonumber \\
\end{eqnarray}

\noindent Note that this contribution has \textit{no explicit
density dependence}. Using Eq.\ (\ref{eqn:fur}), we can write
\begin{eqnarray}
{\cal E}^{(2)}_{\phi-VF} & = &
-\frac{g_{S}^{2}}{2}\int\int\frac{d^{4}k}{(2\pi)^{4}}
\frac{d^{4}q}{(2\pi)^{4}}\,
\Delta_{S}^{0}(k-q) \sum_{m\, =\, 0}^{\infty}\ \sum_{n\, =\, 0}^{\infty}
\left(M^{*}-M\right)^{m+n} \nonumber \\
& & \times \left[\frac{1}{k^{2}+M^{2}}\right]^{m+1}
\left[\frac{1}{q^{2}+M^{2}}\right]^{n+1}
\tr\left[\left(i{\not\!k}-M\right)^{m+1}
\left(i{\not\!q}-M\right)^{n+1}\right] \ ,
\end{eqnarray}

\noindent where $m+n \neq 0$. (The $m+n=0$ term cancels the vacuum
term.) We can use Eq.\ (\ref{eqn:dec}) to solve the trace, or
\begin{eqnarray}
{\cal E}^{(2)}_{\phi-VF} & = &
-g_{S}^{2}\gamma\int\int\frac{d^{4}k}{(2\pi)^{4}}
\frac{d^{4}q}{(2\pi)^{4}}\,
\Delta_{S}^{0}(k-q) \nonumber \\
& & \times \sum_{m\, =\, 0}^{\infty}\ \sum_{n\, =\, 0}^{\infty}
\left(M^{*}-M\right)^{m+n}
\left[\frac{1}{k^{2}+M^{2}}\right]^{m+1}
\left[\frac{1}{q^{2}+M^{2}}\right]^{n+1} \nonumber \\
& & \times \left[g_{m+1}(k^{2},M)g_{n+1}(q^{2},M)
-k\cdot q f_{m+1}(k^{2},M)f_{n+1}(q^{2},M)\right]\ .
\end{eqnarray}

\noindent Now one can expand the baryon denominators around $k^{2}=M^{2}$ and $q^{2}=M^{2}$
and perform a Wick rotation to Euclidean space.
What remains are some integrals over a complicated polynomial in
$k^{2}$, $q^{2}$, $k\cdot q$, $M^{*}$, and $M$. However complicated,
these four-dimensional integrals can be done, and we are left with

\begin{equation}
{\cal E}^{(2)}_{\phi-VF} = \sum_{m\, =\, 1}^{\infty}
b_{m}^{\phi}(g_{S}\phi_{0})^{m}\ , \label{eqn:evf}
\end{equation}

\noindent where $b_{m}^{\phi}$ are constants that depend on $M$. These terms can
also be absorbed into preexisting terms in the lagrangian, and hence
there is no need to calculate them. Fortunately, only a few coefficients are required
for an accurate description of bulk nuclear properties \cite{ref:Fu95}.
Note that the forms in Eqs.\
(\ref{eqn:els}) and (\ref{eqn:evf}) are consistent with the explicit
results for these integrals given in \cite{ref:Fu89}.

\subsection{Long-range Physics}

In this section, we consider the two-loop exchange contributions to
the energy density in the presence of background mean fields.  These
terms are nonanalytic functions of the Fermi momentum and correspond
to nonlocal contributions to the energy. The scalar meson
contribution to the exchange term becomes
\begin{eqnarray}
{\cal E}_{\phi - EX}^{(2)} & = &
-\frac{g_{S}^{2}}{2}\int\int\frac{d^{4}k}{(2\pi)^{4}}\frac{d^{4}q}{(2\pi)^{4}}\,
\Delta_{S}^{0}(k-q) \tr\left[G_{D}^{*}(k)G_{D}^{*}(q)\right] \nonumber \\
& = & {}
\frac{g_{S}^{2}}{2}\int\int\frac{d^{4}k}{(2\pi)^{3}}\frac{d^{4}q}{(2\pi)^{3}}
\frac{\theta(k_{F}-|\vec{k}|)}{2E^{*}(k)}
\frac{\theta(k_{F}-|\vec{q}|)}{2E^{*}(q)} \nonumber \\
& &  \qquad {} \times \delta\left(k_{4}-E^{*}(k)\right)
\delta\left(q_{4}-E^{*}(q)\right)\frac{1}{(k-q)^{2}+m_{S}^{2}} \nonumber \\
& & \qquad {} \times \tr\left[\left(i{\not\! k}-M^{*}\right)
\left(i{\not\! q}-M^{*}\right)\right] \nonumber \\[5pt]
& = & \frac{\gamma g_{S}^{2}}{32\pi^{4}}\int_{0}^{k_{F}}
\frac{|\vec{k}|^{2}d|\vec{k}|}{E^{*}(k)}
\int_{0}^{k_{F}} \frac{|\vec{q}|^{2}d|\vec{q}|}{E^{*}(q)}
\int_{-1}^{1}d(\cos\theta) \nonumber \\
& & \times \left[\frac{E^{*}(k)E^{*}(q)-|\vec{k}||\vec{q}|
\cos\theta+{M^{*}}^{2}}
{2E^{*}(k)E^{*}(q)-2|\vec{k}||\vec{q}|\cos\theta-2{M^{*}}^{2}+m_{S}^{2}}\right]\
, \label{eqn:LLL}
\end{eqnarray}

\noindent where we have used Eq.\ (\ref{eqn:BB3}) and integrated out
most of the angular dependence. The vector meson contribution is
\begin{eqnarray}
{\cal E}_{V - EX}^{(2)} & = &
\frac{g_{V}^{2}}{2}\int\int\frac{d^{4}k}{(2\pi)^{4}}\frac{d^{4}q}{(2\pi)^{4}}\,
{\cal D}_{\mu\nu}^{0}(k-q)
\tr\left[\gamma_{\mu}G_{D}^{*}(k)\gamma_{\nu}G_{D}^{*}(q)\right] \nonumber \\
& = &
-\frac{g_{V}^{2}}{2}\int\int\frac{d^{4}k}{(2\pi)^{3}}\frac{d^{4}q}{(2\pi)^{3}}
\frac{\theta(k_{F}-|\vec{k}|)}{2E^{*}(k)}
\frac{\theta(k_{F}-|\vec{q}|)}{2E^{*}(q)} \nonumber \\
& & \qquad {} \times
\delta\left(k_{4}-E^{*}(k)\right)\delta\left(q_{4}-E^{*}(q)\right)
\frac{1}{(k-q)^{2}+m_{V}^{2}}\, \delta_{\mu\nu} \nonumber \\
& & \qquad {} \times \tr\left[\gamma_{\mu}\left(i{\not\!
k}-M^{*}\right)
\gamma_{\nu}\left(i{\not\! q}-M^{*}\right)\right] \nonumber \\[5pt]
& = & \frac{\gamma g_{V}^{2}}{16\pi^{4}}\int_{0}^{k_{F}}
\frac{|\vec{k}|^{2}d|\vec{k}|}{E^{*}(k)}\int_{0}^{k_{F}}
\frac{|\vec{q}|^{2}d|\vec{q}|}{E^{*}(q)}
\int_{-1}^{1}d(\cos\theta) \nonumber \\
& & \times \left[\frac{E^{*}(k)E^{*}(q)-|\vec{k}||\vec{q}|
\cos\theta-2{M^{*}}^{2}}
{2E^{*}(k)E^{*}(q)-2|\vec{k}||\vec{q}|\cos\theta-2{M^{*}}^{2}+m_{V}^{2}}\right]\
,
\end{eqnarray}

\noindent where we integrated out most of the angular dependence.
The contribution to the integral from the longitudinal portion of the
vector propagator vanishes if one works out the trace.
Finally, we consider the two-loop contribution arising from pion
exchange. We remove most of the angular dependence and arrive at the
result
\begin{eqnarray}
{\cal E}_{\pi - EX}^{(2)} & = &
\frac{g_{A}^{2}}{2f_{\pi}^{2}}
\int\int\frac{d^{4}k}{(2\pi)^{4}}\frac{d^{4}q}{(2\pi)^{4}}\,
\Delta_{\pi}^{ab}(k-q)  \nonumber \\
& & \times  \tr\left[(\not\! k - \not\! q)\gamma_{5}\frac{\tau_{a}}{2}
G^{*}_{D}(k) (\not\! k - \not\! q)\gamma_{5}\frac{\tau_{b}}{2}G^{*}_{D}(q)\right]
\vphantom{\int}  \nonumber \\
& = & -\frac{g_{A}^{2}}{2f_{\pi}^{2}}
\int\int\frac{d^{4}k}{(2\pi)^{3}}\frac{d^{4}q}{(2\pi)^{3}}
\frac{\theta(k_{F}-|\vec{k}|)}{2E^{*}(k)}
\frac{\theta(k_{F}-|\vec{q}|)}{2E^{*}(q)} \nonumber \\
& & \qquad {} \times \delta\left(k_{4}-E^{*}(k)\right)
\delta\left(q_{4}-E^{*}(q)\right)\frac{1}{(k-q)^{2}+m_{\pi}^{2}}\delta_{ab} \nonumber \\
& & \qquad {} \times \tr\left[(\not\! k - \not\!
q)\gamma_{5}\frac{\tau_{a}}{2}\left(i{\not\! k}-M^{*}\right)
(\not\! k - \not\! q)\gamma_{5}\frac{\tau_{b}}{2}\left(i{\not\! q}-M^{*}\right)\right]
\nonumber \\[5pt]
& = & \frac{\gamma g_{A}^{2}}{8\pi^{4}f_{\pi}^{2}}
\left(\frac{5\gamma-8}{16}\right){M^{*}}^{2}
\int_{0}^{k_{F}}\frac{|\vec{k}|^{2}d|\vec{k}|}{E^{*}(k)}
\int_{0}^{k_{F}}\frac{|\vec{q}|^{2}d|\vec{q}|}{E^{*}(q)}
\int_{-1}^{1}d(\cos\theta) \nonumber \\
& & \times \left[\frac{E^{*}(k)E^{*}(q)-|\vec{k}||\vec{q}|
\cos\theta-{M^{*}}^{2}} {2E^{*}(k)E^{*}(q)-2|\vec{k}||\vec{q}|
\cos\theta-2{M^{*}}^{2}+m_{\pi}^{2}}\right]\ .
\end{eqnarray}

\section{Discussion}

In this work, we performed the loop expansion to the two-loop level
for a model QHD lagrangian. This expansion provided a simple scheme
for separating the short- and long-range dynamics order by order and
for analyzing their structures. The short-range, local physics (including
all the divergences) was isolated and absorbed into the coefficients
of the lagrangian. The long-range dynamics is nonlocal and has to be
calculated explicitly. For the purposes of this work, we are
interested in applying the loop expansion only to ordinary,
symmetric nuclear matter. For that reason, we neglected both
electromagnetic effects and $\rho$ meson exchange, although they
both appear in the mean field lagrangian \cite{ref:Fu97,ref:Se97}.
In addition, we exclude the tensor terms in the fermion sector and
the nonlinear meson self-couplings. The exploration of the effects
of these terms at the two-loop level is left for future work. The
inclusion of the two-loop effects also calls into question whether
the naturalness assumption \cite{ref:Fu97,ref:Ma84,ref:Ge93} is
preserved; we partially answer that question here, but a more complete proof
that naturalness holds is also left for a subsequent investigation.

\begin{figure}[t!]
\begin{center}
\includegraphics[width=4 in]{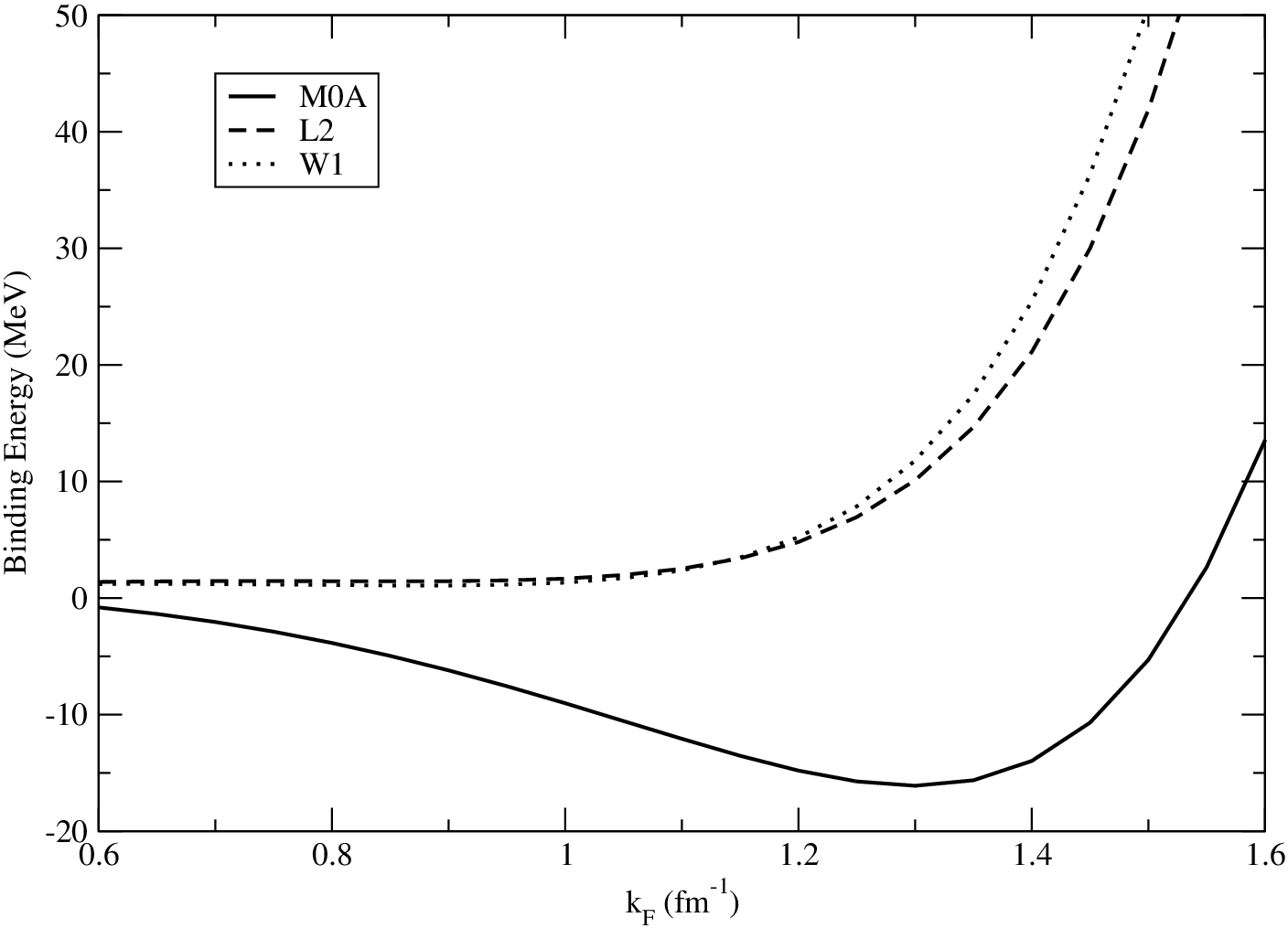}
\caption{Comparison of the nuclear binding curves for the sets L2
and W1 (both with one-loop parameters but with the two-loop
contributions included) and M0A (two-loop level).\vspace{.1 in}} \label{fig:loop2}
\end{center}
\end{figure}

\begin{figure}
\begin{center}
\includegraphics[width=4 in]{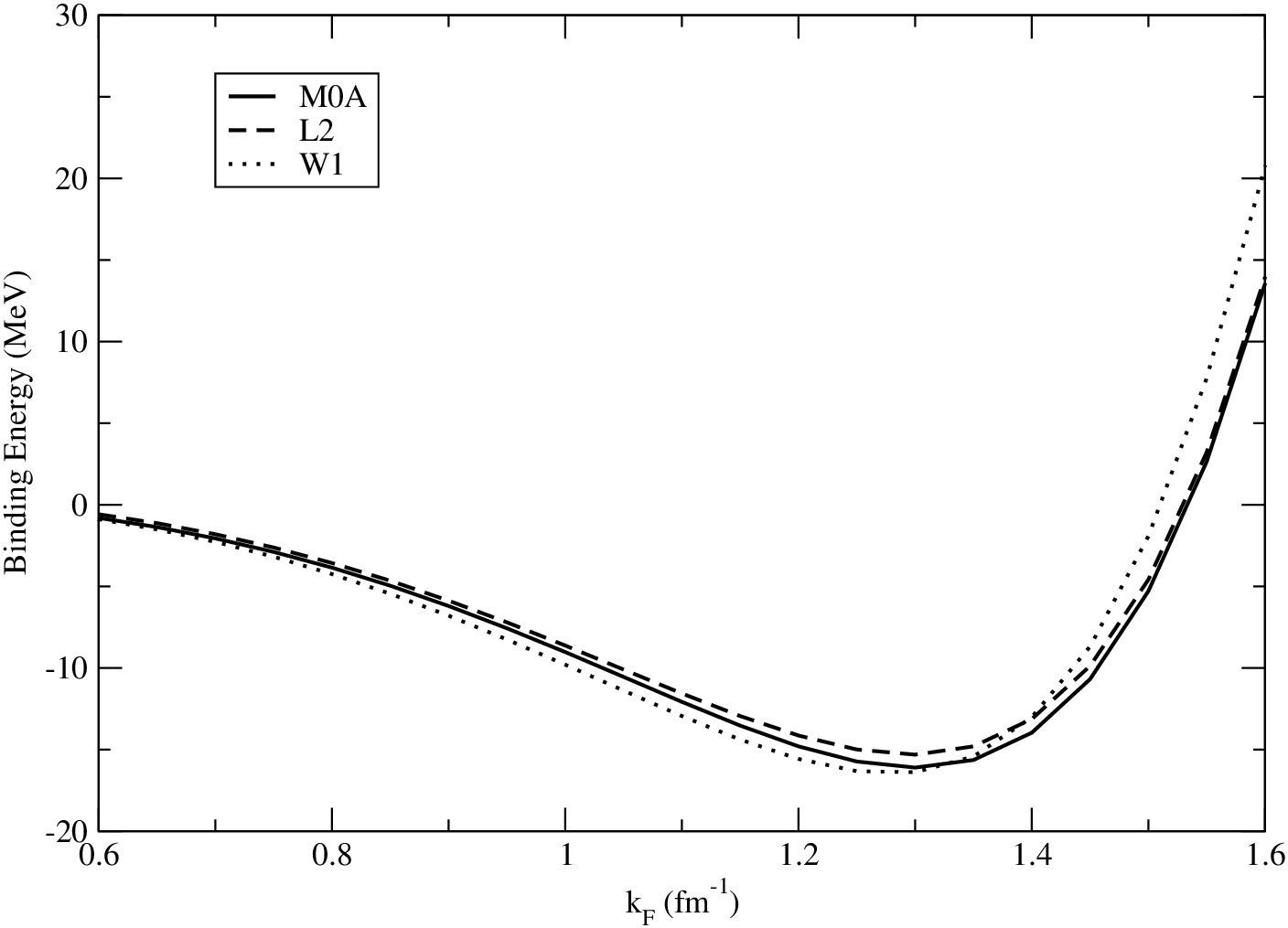}
\caption{Comparison of the nuclear binding curves for the sets L2
and W1 (both one-loop level) and M0A (two-loop level).\vspace{.2 in}}
\label{fig:loop3}
\end{center}
\end{figure}

Now we consider the numerical analysis of the three surviving
two-loop integrals (the exchange terms). We use the parameter sets
listed in Table \ref{tab:1}. For the pion term, we use
$g_{A}^{2}=1.5876$ and $f_{\pi}=93$ MeV \cite{ref:Fu97}. In
addition, both L2 and W1 sets lead to equilibrium at $k_{\mathrm
F}=1.3$ fm$^{-1}$. The mean meson fields are determined by
extremizing the meson field equations and are used as input to
the exchange integrals. Then the full two-loop energy density is
extremized with respect to the meson fields.
The results for all three exchange terms are
shown in Table \ref{tab:2} for both L2 and W1 parameter sets. The
third set (M0A) listed in Table \ref{tab:1} was fit at the two-loop
level to nuclear equilibrium (${\cal E}/\rho_{B}-M=-16.10$ MeV and
$k_{\mathrm F}=1.3$ fm$^{-1}$) by adjusting $g_S$ and $g_V$ using a
downhill simplex method to minimize a least-squares fit (with
respective weights of $0.0015$ and $0.002$). The binding curves of
all three sets are shown in Fig.\ \ref{fig:loop2}. The curves for
the L2 and W1 sets include the two-loop contributions and are
compared to the saturation curve of the M0A set at the two-loop
level. Fig.\ \ref{fig:loop3} shows results when the L2 and W1 sets
are evaluated at the one-loop level, and the M0A set is evaluated at
the two-loop level. First, we note that nuclear saturation can be
reproduced at the two-loop level with a parameter set (M0A) that is
natural $(g_{S,V} \approx 4\pi)$. Second, Fig.\ \ref{fig:loop2}
shows that while the two-loop contributions are not large, they are
not negligible either.

\begin{table}
\begin{center}
\begin{tabular}{|c|c|c|c|} \hline
             & L2 \cite{ref:Se97} & W1 \cite{ref:Fu97} & M0A     \\
\hline $m_{S}/M$    & 0.55378            & 0.60305            &0.54     \\
\hline $m_{V}/M$    & 0.83387            & 0.83280 & 0.83280 \\
\hline $g_{S}/4\pi$ & 0.83321            & 0.93797 & 0.79361 \\
\hline $g_{V}/4\pi$ & 1.09814            & 1.13652 & 0.96811 \\
\hline
\end{tabular}
\caption{Parameter sets used in this work.} \label{tab:1}
\end{center}
\end{table}
\begin{table}
\begin{center}
\begin{tabular}{|c|c|c|c|} \hline
                      & L2 \cite{ref:Se97} & W1 \cite{ref:Fu97} & M0A    \\
\hline ${\cal E}_{\phi-EX}^{(2)}\vphantom{\displaystyle\sum}$ & $\zz 42.59$ & $\zz 46.65$ & $\zz 40.30$ \\
\hline ${\cal E}_{V-EX}^{(2)} \vphantom{\displaystyle\sum} $  & $-29.51$    & $-30.60$ & $-23.14$ \\
\hline ${\cal E}_{\pi-EX}^{(2)} \vphantom{\displaystyle\sum}$ & $\zz 12.39$ & $\zz 12.21$ & $\zz 12.44$ \\
\hline
\end{tabular}
\caption{Size of two-loop integrals for the parameter sets in Table
\ref{tab:1}. Values are in MeV.\vspace{.1 in}} \label{tab:2}
\end{center}
\end{table}


Next, we compare the magnitudes of these integrals to terms in the
meson sector at the mean field level. However, neither set L2 nor W1
includes nonlinearities in the meson fields. Therefore, we will
compare the exchange integrals to terms from the sets Q1 and Q2
\cite{ref:Fu97}. Here Q1 and Q2 both include cubic and quartic
scalar field self-couplings and Q2 also contains a quartic vector
field self-coupling. These nonlinear terms are contained in 
\begin{equation}
{\cal L}_{NL}=-\left( { {{\kappa}_3 } \over {3!} } { {g_S \phi} \over M } +
{ {{\kappa}_4 } \over {4!} } { {g^2_S {\phi}^2} \over {M^2} }\right)
m^2_S {\phi}^2 + { 1 \over {4!} } \,{\zeta}_0 g_V^2 {\left( V_{\mu}
V^{\mu} \right)}^2
\end{equation}

\noindent of Eq.\ (\ref{eqn:lagrangian}). This does not imply that
these nonlinearities were taken into account when the two-loop
integrals were calculated; it is simply instructive to compare the
relative sizes of the mean field nonlinearities and the two-loop
contributions. Note that the Q1 and Q2 mean fields were obtained by 
extremizing the one-loop energy, including ${\cal L}_{NL}$,
while the $M^{*}$ in the exchange integrals was obtained by minimizing 
${\cal E}^{(2)}$ of Eq.\ (\ref{eqn:full}).
While the inclusion of the nonlinear terms in the
meson sector at the two-loop level has yet to be performed, a
cursory glance tells us that they will affect only the meson
propagators in two-loop integrals. Thus the overall magnitude of
these terms is not expected to change much when these nonlinear
effects are included.

\begin{figure}
\begin{center}
\includegraphics[width=4 in]{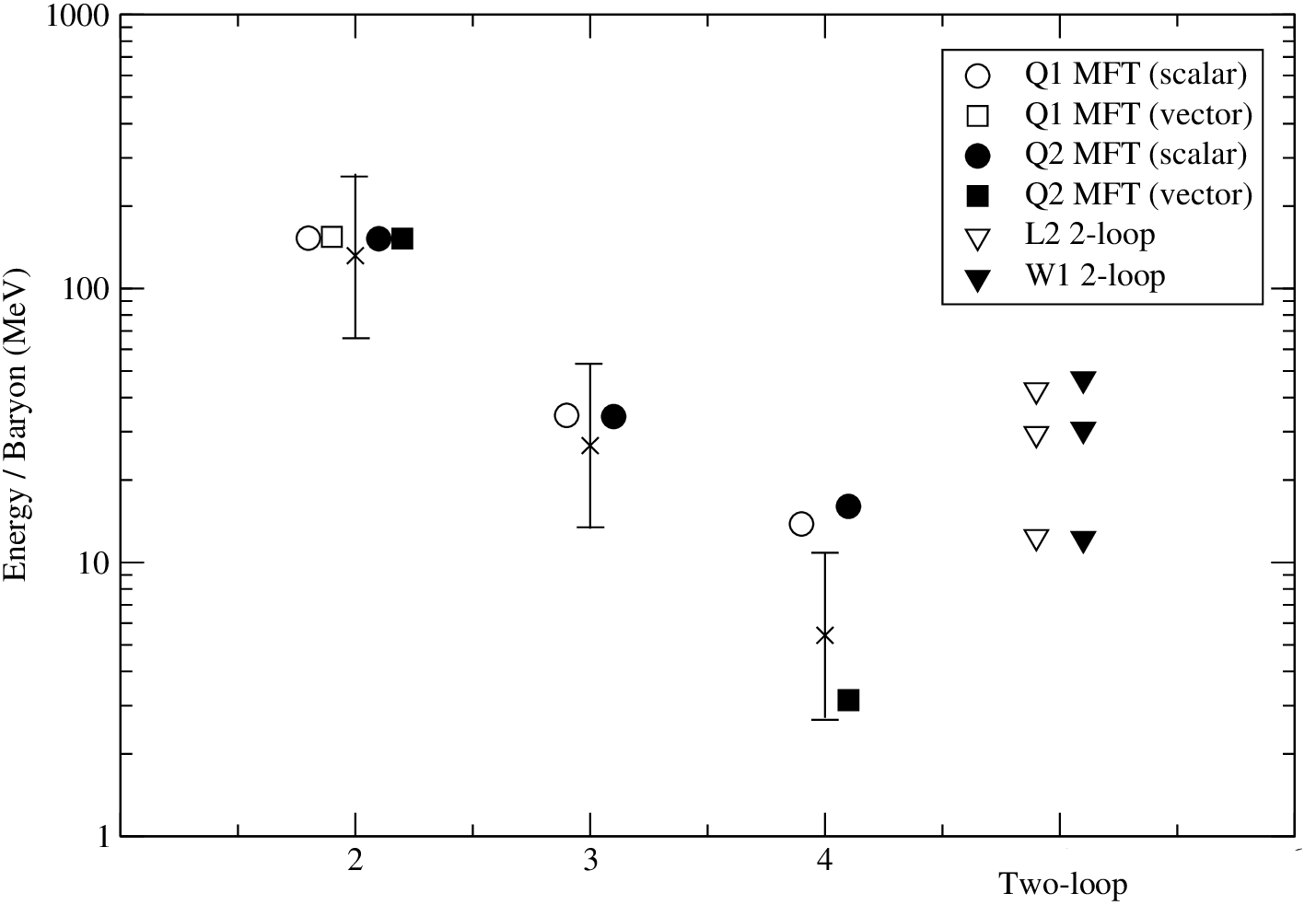}
\caption{Comparison of the magnitudes of the mean field terms in the
meson sector with the two-loop exchange integrals. The inverted
triangles represent, from top to bottom, the scalar, vector, and pion
two-loop integrals. The abscissa
represents the order $\nu$ in the power counting at the mean field
level \cite{ref:Fu97}. Absolute values are shown.} \label{fig:loop1}
\end{center}
\end{figure}

Fig.\ \ref{fig:loop1} shows the results of this comparison.  The
crosses in Fig.\ \ref{fig:loop1} represent the expected magnitude
per order of terms in the meson sector using the rules of naive
dimensional analysis (NDA) \cite{ref:Fu97} with the chiral symmetry breaking
scale of $\Lambda=650$ MeV. Thus, one observes that
the two-loop integrals are roughly equivalent to $\nu = 3$ order in
the power counting in the mean field lagrangian. While this is not
large, they cannot be neglected in a description of nuclear matter
saturation properties, particularly in view of the nearly complete
cancellation of scalar and vector terms at order $\nu = 2$. We
emphasize that the explicit computation of short-range, two-loop
contributions leads to unnaturally large terms in
the energy of nuclear matter \cite{ref:Fu89}; in the EFT approach, these
contributions are absorbed in the local counterterms and are either
removed by field redefinitions or are determined by fitting to
empirical nuclear data.

We stress that the results shown here are for the crudest truncation in the underlying
lagrangian. It is well known that an accurate description of nuclear saturation
(and the equation of state) requires the inclusion of nonlinear meson self-interactions,
which is the subject of future work \cite{ref:Mc07}. Furthermore, the question of whether
or not the loop expansion is valid for QHD cannot be answered at 
the two-loop level. While two-loop calculations are a necessary step in that direction, an
investigation into higher-loop effects is required; this is also the subject of
future work \cite{ref:Mc07b}.

\subsection{Infrared Regularization}

The separation of the short-range and long-range dynamics, and the
subsequent absorption of the short-range physics into the lagrangian
are analogous to Infrared Regularization in Chiral Perturbation Theory
\cite{ref:Ta96,ref:El98,ref:Be99,ref:Be00,ref:Sc03}. In the language
of Ellis and Tang \cite{ref:Ta96,ref:El98}, a loop integral can be
separated into hard and soft components
\begin{equation}
G=(G-\hat{R}\hat{S}G)+\hat{R}\hat{S}G\ ,
\end{equation}

\noindent where $G$ is the loop integral, $\hat{S}$ projects out the
soft part, and $\hat{R}$ renormalizes $\hat{S}G$ to remove the
ultraviolet divergences. To extract the soft part, do the
following: take the loop momentum to be of order $Q$, make a $Q/M$
expansion of the integrand, and interchange the order of integration
and summation. Now consider a loop integral with momentum $q$,
specifically the $q_4$ part of the integral. Closing the contour by
a semicircle at infinity, we get the sum of three contributions: the
semicircle, the soft poles, and the hard poles. The soft and hard
poles are of order $Q$ and $M$ respectively. (It has been assumed
that the hard and soft poles can be separated, as is the case in
theories with Goldstone bosons and massive baryons.) The semicircle
may produce divergences, but these can be removed by the usual
renormalization. The soft poles cannot be expanded in $Q/M$, such as
those in the pion propagator. However, a $Q/M$ expansion can be made
around the hard poles since the loop momentum is of order $Q$.
Finally, integrating term by term removes the hard contribution, and
thus we are left with the unrenormalized soft part, in which
ultraviolet divergences may still occur (which are removed by
applying $\hat{R}$). Alternative implementations can be found in
\cite{ref:Be99,ref:Be00,ref:Sc03}.

Since the prescription for $\hat{R}\hat{S}G$ contains all the soft
parts, the hard portion must be given by $(G-\hat{R}\hat{S}G)$; the
hard part involves only the large momentum contributions, including
some ultraviolet divergences. We can write the hard part as a series
of local counterterms---a well-known result that is fundamental to
the idea of effective field theories. Indeed, large momenta
correspond to short distances that are tiny compared with the
wavelengths of the external particles, so the effects can be described
by a local interaction. Thus, we can perform the \textit{extra renormalization} of
absorbing the hard parts into the parameters of the lagrangian.

In our case, we separate the baryon propagator into the Feynman and
Density parts. The Feynman part contains the high-momentum
contributions and also contains all the hard poles. The Density part contains all
the soft poles describing the valence nucleons, which occur at lower
momenta. (Note that the momenta in the meson propagators are always spacelike.) One
can write the one-loop self-energy for the scalar two-loop integral
as
\begin{eqnarray}
\Sigma^{*\phi}(k) & = & -\frac{g_{S}^{2}}{2}\int\frac{d^{4}q}{(2\pi)^{4}}
\,\Delta_{S}^{0}(k-q)G^{*}(q) \nonumber \\
& = & -\frac{g_{S}^{2}}{2}\int\frac{d^{4}q}{(2\pi)^{4}}
\,\Delta_{S}^{0}(k-q)G^{*}_{F}(q) -
\frac{g_{S}^{2}}{2}\int\frac{d^{4}q}{(2\pi)^{4}}
\,\Delta_{S}^{0}(k-q)G^{*}_{D}(q)\ .
\end{eqnarray}

\noindent The first term contains only the hard poles and can be
absorbed by the processes outlined above: we expand either in powers
of momentum or in powers of the mean fields (Furry's theorem). Then
interchange orders of summation and integration; what's left are
series in the mean fields.  The second term contains the soft poles,
but they are regularized by the theta function in the Density
propagator and hence this contribution is finite. As a result, the
procedure outlined in this work is analogous to Infrared
Regularization.

\subsection{Power Counting}

To illustrate the order of the two-loop exchange integrals, we expand them
in powers of momenta and pick out the dominant contributions. We can
expand the isoscalar meson propagators as
\begin{equation}
\frac{1}{x+1}=1-x+x^{2}-x^{3}+\ldots \ ,
\label{eqn:exp1}
\end{equation}

\noindent where $x=(k-q)/m_{S}\leq 1$ for the scalar meson and
$x=(k-q)/m_{V}\leq 1$ for the vector meson. This is essentially an
expansion of meson exchange into contact and gradient interaction
terms.  Thus, one can pick out the leading (i.e., contact) terms
\begin{eqnarray}
{\cal E}_{\phi-EX}^{(2)} & \approx &
\frac{g_{S}^{2}}{4\pi^{4}m_{S}^{2}}\int_{0}^{k_{F}}
\frac{|\vec{k}|^{2}d|\vec{k}|}{E^{*}(k)}
\int_{0}^{k_{F}}\frac{|\vec{q}|^{2}d|\vec{q}|}{E^{*}(q)}
\left(E^{*}(k)E^{*}(q)+{M^{*}}^{2}\right) \nonumber \\
& = & \frac{g_{S}^{2}}{16m_{S}^{2}}\left(\rho_{B}^{2}+\rho_{S}^{2}\right)
\label{eqn:sc_l} \\
{\cal E}_{V-EX}^{(2)} & \approx &
\frac{g_{V}^{2}}{2\pi^{4}m_{V}^{2}}\int_{0}^{k_{F}}
\frac{|\vec{k}|^{2}d|\vec{k}|}{E^{*}(k)}
\int_{0}^{k_{F}}\frac{|\vec{q}|^{2}d|\vec{q}|}{E^{*}(q)}
\left(E^{*}(k)E^{*}(q)-2{M^{*}}^{2}\right) \nonumber \\
& = &
\frac{g_{V}^{2}}{8m_{V}^{2}}\left(\rho_{B}^{2}-2\rho_{S}^{2}\right)\
 \label{eqn:vec_l}
\end{eqnarray}

\noindent In the case of the pion two-loop exchange integral, the
pion mass is too small for Eq.\ (\ref{eqn:exp1}) to converge.
Therefore, we instead take the chiral limit ($m_{\pi}\rightarrow 0$)
to get the dominant term
\begin{equation}
{\cal E}_{\pi-EX}^{(2)} \approx
\frac{3g_{A}^{2}}{32f_{\pi}^{2}}\rho_{S}^{2} + O( m_{\pi}^2 ) \ .
\label{eqn:pion_l}
\end{equation}

\noindent These are the terms which set the scale for the exchange
contributions.

One can acquire the scale by comparison to terms at the
mean field level; for instance, the terms in Eqs.\ (\ref{eqn:sc_l}) and (\ref{eqn:vec_l})
in ratio with the leading vector mean field term are (where we have
used the vector meson field equation)
\begin{eqnarray}
\frac{\displaystyle{\frac{g_{S}^{2}}{16m_{S}^{2}}}
\left(\rho_{B}^{2}+\rho_{S}^{2}\right)}
{\displaystyle{\frac{g_{V}^{2}}{2m_{V}^{2}}}\rho_{B}^{2}} & \approx & \frac{2}{7} \\
\frac{\displaystyle{\frac{g_{V}^{2}}{8m_{V}^{2}}}
\left(\rho_{B}^{2}-2\rho_{S}^{2}\right)}
{\displaystyle{\frac{g_{V}^{2}}{2m_{V}^{2}}}\rho_{B}^{2}} & \approx & -\frac{1}{5} \ .
\end{eqnarray}

\noindent Next, we consider the leading terms at normal density
($k_{\mathrm F}=1.3$ fm$^{-1}$) using W1 \cite{ref:Fu97}; we get
${\cal E}_{\phi-EX}^{(2)}/\rho_{B} \approx 57$ MeV, ${\cal
E}_{V-EX}^{(2)}/\rho_{B} \approx -34$ MeV, and ${\cal
E}_{\pi-EX}^{(2)}/\rho_{B} \approx 17$ MeV. These terms are roughly
third order $(\nu = 3)$ in the power counting. Observe that the
self-energies are reproduced at the 10--40\% level using only the
leading-order contact term or with the chiral limit.  To incorporate
the corrections to these results, it is clearly most efficient to
retain the full meson propagators in the exchange integrals.

Exchange interactions are two-body, so they go like $O(\rho^{2})$ in
$\cal E$. This is the same dependence as the two-body mean field
terms at $O(\nu=2)$. Because of the lack of a spin--isospin sum in
the exchange terms, the exchange terms are numerically smaller and
contribute at the same level as the $O(\nu=3)$ terms. A deeper
understanding of how loops fit into the finite-density power-counting 
scheme will require a look at higher-order terms in the
loop expansion, which is currently under investigation.
\begin{figure}
\begin{center}
\includegraphics[width=4 in]{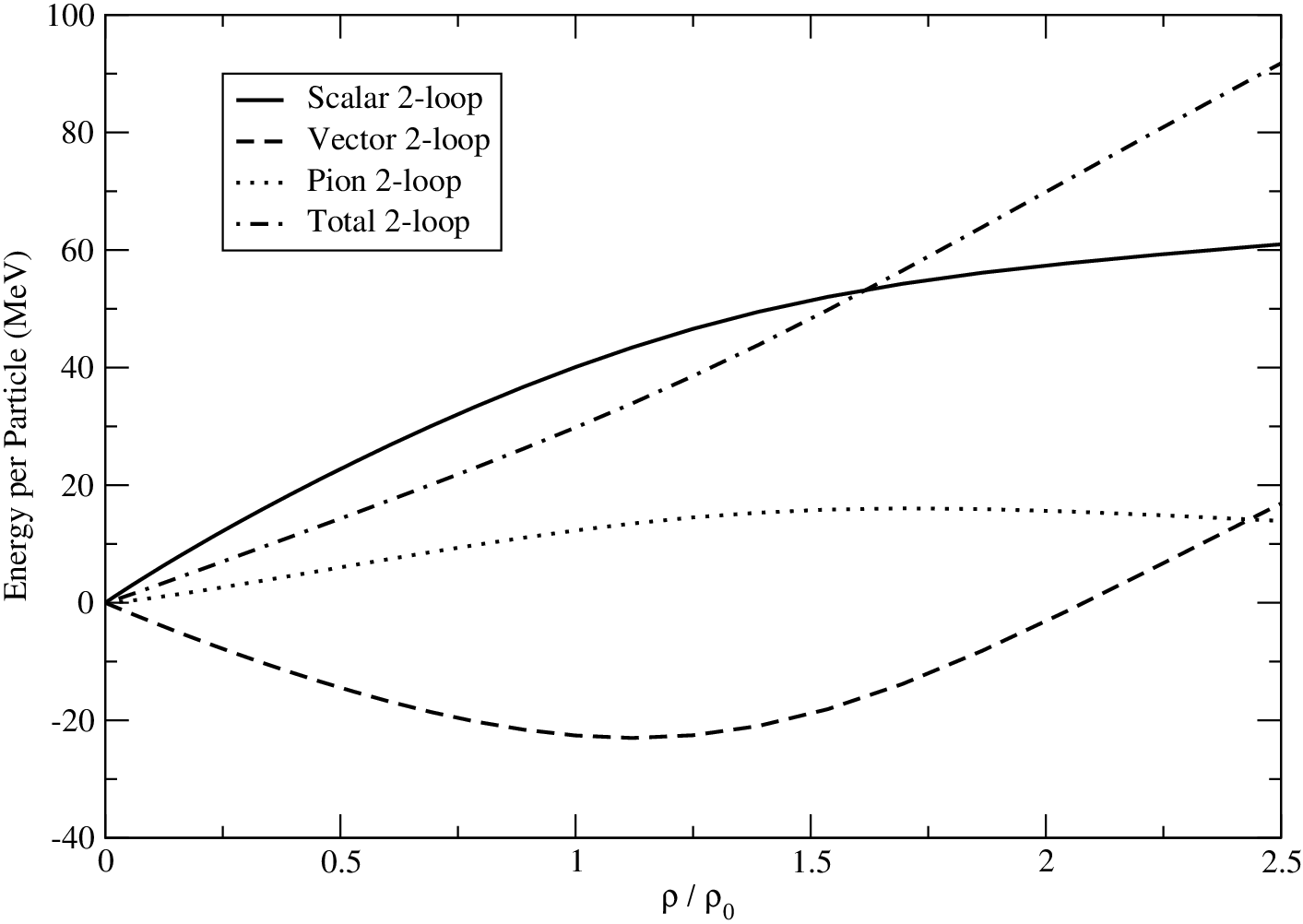}
\caption{The separate and total two-loop contributions to the energy 
per particle versus the density.}
\label{fig:loop4}
\end{center}
\end{figure}

In Fig.\ \ref{fig:loop4}, we show the individual and total two-loop
contributions to the energy per particle plotted as a function of
the density $\rho=\rho_{B}$. Since the ratio $\rho_{S}/\rho_{B}$ decreases as
$\rho_{B}$ increases, one sees that the qualitative behavior of the
scalar and vector contributions agrees with the leading-order estimates
in Eq.\ (\ref{eqn:sc_l}) and (\ref{eqn:vec_l}). This is not surprising, since the 
two-loop exchange integrals are dominated by the contact terms, which have a simple
density dependence. The total two-loop contribution is almost linearly
dependent on the density, which occurs ``by construction'' with the appropriately
refit coupling parameters. While this implies that the two-loop terms are
relatively short-ranged, they are nonetheless nonlocal and hence they cannot be
absorbed directly into the parameterization via Infrared Regularization.

\section{Summary}

In this work, we studied two-loop corrections to symmetric nuclear
matter in a covariant effective field theory. The loop expansion
gives a straightforward way to separate the short-distance physics
from the long-distance physics.  The former can be absorbed into
counterterms already present in the effective lagrangian, and they
are either removed by field redefinitions or fitted to empirical
data.  The remaining long-range exchange integrals are nonlocal and
must be computed explicitly.  They produce modest corrections to the
nuclear binding curve and can be compensated by a small adjustment
of the coupling parameters.

Since exchange integrals in effective hadronic field theories have
been studied for more than 30 years \cite{ref:Ch77}, it is important
to enumerate the new features of our calculations.  First, the QHD
model studied here has its basis in a Lorentz-covariant,
chiral-invariant, hadronic effective field theory that is tailored
to the nuclear many-body problem \cite{ref:Fu97} and that
successfully describes bulk and single-particle nuclear properties
at the one-loop level
\cite{ref:Fu97,ref:Se97,ref:Fu99,ref:Fu03,ref:Se04,ref:Fu04,ref:Mc02,ref:Mc04,ref:Mc05,ref:He02}.
Second, using standard procedures of EFT, the loop expansion at
\textit{finite density} provides a systematic, well-defined
treatment of the short- and long-range contributions to the
integrals that can be extended to higher orders in loops
\cite{ref:Hu00,ref:Mc07}. Third, when interpreted in the context of
density functional theory, the exchange contributions to the energy
introduce nonanalytic density dependence that is qualitatively
different from that appearing in the mean-field theory.  This should
allow for an improved approximation to and parametrization of the
exact energy functional. In addition, as a criterion for discussing
the size of the exchange contributions, one can readjust the
coupling parameters to reproduce the nuclear matter equilibrium
point and see if they remain of natural size; this was indeed the
case with the two couplings adjusted here. The naturalness of the
remaining parameters in the underlying EFT will be studied in future
work. Finally, the size of the two-loop integrals (roughly third
order in the mean field power counting) was determined. Fuller
consideration of how the loop expansion fits into the power counting
is left for future investigation.

\section*{Acknowledgements}

We thank our colleagues R. J. Furnstahl and J. D. Walecka for valuable
comments on the manuscript. This work was supported in part by the Department
of Energy under Contract No.\ DE--FG02--87ER40365.

\end{document}